\documentclass[%
nofootinbib,
 aip,
 amsmath,amssymb,
 reprint,%
]{revtex4-1}

\usepackage{graphicx}
\usepackage{dcolumn}
\usepackage{bm}

\usepackage[utf8]{inputenc}
\usepackage[T1]{fontenc}
\usepackage{mathptmx}
\usepackage{etoolbox}
\usepackage{overpic}
\usepackage{tikz}

\usepackage[hidelinks]{hyperref}
\usepackage[nameinlink,capitalize]{cleveref}

\usepackage{xcolor}
\newcommand{\vp}{\varphi}

\newcommand{\half}{\frac{1}{2}}
\renewcommand{\d}{\,\text{d}}

\makeatletter
\def\@email#1#2{%
 \endgroup
 \patchcmd{\titleblock@produce}
  {\frontmatter@RRAPformat}
  {\frontmatter@RRAPformat{\produce@RRAP{*#1\href{mailto:#2}{#2}}}\frontmatter@RRAPformat}
  {}{}
}%
\makeatother
\begin{document}

\preprint{AIP/123-QED}

\title[Continuum limit of the adaptive Kuramoto model]{Continuum limit of the adaptive Kuramoto model}
\author{Rok Cestnik}
 \email{rok.cestnik@math.lth.se}
\author{Erik A. Martens}%
 \email{erik.martens@math.lth.se}
\affiliation{ 
Centre for Mathematical Science, Lund University, Sölvegatan 18, 22100, Lund, Sweden
}%

\date{\today}

\begin{abstract}
We investigate the dynamics of the adaptive Kuramoto model with slow adaptation in the continuum limit, $N\to\infty$. This model is distinguished by dense multistability, where multiple states coexist for the same system parameters. The underlying cause of this multistability is that some oscillators can lock at different phases or switch between locking and drifting depending on their initial conditions. We identify new states, such as two-cluster states. To simplify the analysis we introduce an approximate reduction of the model via row-averaging of the coupling matrix. We derive a self-consistency equation for the reduced model and present a stability diagram illustrating the effects of positive and negative adaptation. Our theoretical findings are validated through numerical simulations of a large finite system. Comparisons to previous work highlight the significant influence of adaptation on synchronization behavior.
\end{abstract}

\maketitle

\begin{quotation}
The synchronization of coupled oscillators is a fundamental phenomenon observed in many natural and technological systems, from heartbeats to power grids and neural networks. Traditional models like the Kuramoto model have helped us understand how synchronization emerges, but real-world systems often involve adaptive couplings that evolve in time in dependence of the dynamics occurring on the network nodes, i.e., the oscillators. The resulting dynamics \emph{on} and \emph{of} the network leads to \emph{co-evolutionary network dynamics}, posing a mathematically challenging situation to analyze. Our study explores the adaptive Kuramoto model for a large number of oscillators (the continuum limit), focusing on how slow adaptation influences synchronization. We identify new states, such as two-cluster states, and introduce a simplified model to analyze large systems easier. Our findings reveal that adaptation significantly affects synchronization behavior, leading to multiple coexisting states for the same parameters and complex dynamics depending on initial conditions. These theoretical insights are supported by numerical simulations, providing a deeper understanding of the role of adaptation in synchronized systems.
\end{quotation}

\section{\label{sec:intro}Introduction}
The synchronization of coupled oscillators is a fundamental phenomenon observed in various natural and technological systems~\cite{synch_book,strogatz_synch}, such as heartbeats~\cite{heart_example}, power grids~\cite{power_grids_example}, neural networks~\cite{neural_networks_example1, neural_networks_example2} etc. Traditional models like the Kuramoto model have significantly advanced our understanding of how synchronization emerges. However, real-world systems often exhibit adaptive coupling, which can enhance synchronization and substantially broadens the scope of dynamic behaviors.

Adaptive networks, where coupling weights evolve based on the dynamics of the nodes, are prevalent in nature and technology~\cite{adaptive_example1,adaptive_example2}. This adaptation can lead to complex behaviors not seen in static networks, necessitating further study. Systems like the vascular network~\cite{vascular_example}, the glymphatic system of the brain~\cite{glymphatic_example1}, osteocyte networks~\cite{osteocyte_example}, social networks~\cite{social_example}, and particularly neural networks~\cite{neural_net_example} demonstrate the importance of such adaptive mechanisms for their efficient functioning.

In this context, studying large systems is important, as many natural systems consist of a vast number of interacting components. By examining the continuum limit, where the number of oscillators approaches infinity, we can gain deeper insights into the emergent behaviors and fundamental properties of large networks, benefiting from the simplifications that this theoretical limit provides. Thus, we also generalize a previous study on the adaptive Kuramoto model carried out for  two oscillators~\cite{juttner_martens_2023}.

In this paper, we explore the dynamics of the adaptive Kuramoto model in the continuum limit with slow adaptation. Section II examines the full equations of a large finite system through numerical simulations, identifying basic stationary and non-stationary states. Sec.~III introduces an approximate model that reduces the dimensionality by considering the dynamics of row-averages of the coupling matrix. The reduced model facilitates analysis and is computationally more cost effective to simulate for large systems -- this is the model the subsequent analysis is based on. Sec.~IV introduces the notion of locked and drifting oscillators which we use in the following Sec.~V where a self-consistency equation for the order parameter is derived. Sec.~VI analytically examines the transition to synchrony while Sec.~VII presents a detailed stability diagram based on the analysis of the self-consistency equation. Sec.~VIII compares the results of the self-consistency with numerical simulations of the finite system. 

\section{Microscopic model}\label{sec:full_system}
We generalize the Kuramoto mode to include adaptive rules, similar to those explored in Refs.~\cite{adaptive_example1,aoki_aoyagi_2009}, where the system consists of $N$ phases $\vp_i$ and $N^2$ edge variables $\kappa_{ij}$ evolving according to the following equations~\cite{juttner_martens_2023}
\begin{subequations}
    \begin{align}
    \dot{\vp}_i &= \omega_i + \frac{1}{N}\sum_{j=1}^N \kappa_{ij} \sin(\vp_j-\vp_i) \label{eq:full_system_fi}\,,\\
    \dot{\kappa}_{ij} &= \epsilon(1+a\cos(\vp_j-\vp_i)-\kappa_{ij}) \label{eq:full_system_k}\,. 
    \end{align}
\label{eq:full_system}
\end{subequations}
The concept behind this choice of the adaptation rule~\cite{juttner_martens_2023} is that as the adaptation parameter $a$ goes to zero, all edges asymptotically go to 1 and the original Kuramoto model is recovered. 
The adaptation rule~\eqref{eq:full_system_k} is such that the coupling strength between two oscillators tends to increase if the oscillators are in sync, and decrease if they are in an anti-sync configuration~\cite{adaptive_example1}. For oscillators that are uncorrelated, the coupling between them tends to one. 
Throughout this paper we consider large networks near the continuum limit $N\to\infty$ with slow adaptation $\epsilon \ll 1$.  

We consider heterogeneous frequencies $\omega_i$ sampled from a normal distribution $g(\omega) = \frac{1}{\sigma}\frac{1}{\sqrt{2\pi}} \exp\left(-\frac{1}{2}\frac{\omega^2}{\sigma^2}\right)$. The main parameters we are going to vary are the frequency distribution width $\sigma$ and the adaptivity $a$. 

\subsection{Numerical simulations}
\begin{figure}[!h]
\begin{overpic}[width=0.45\textwidth]{"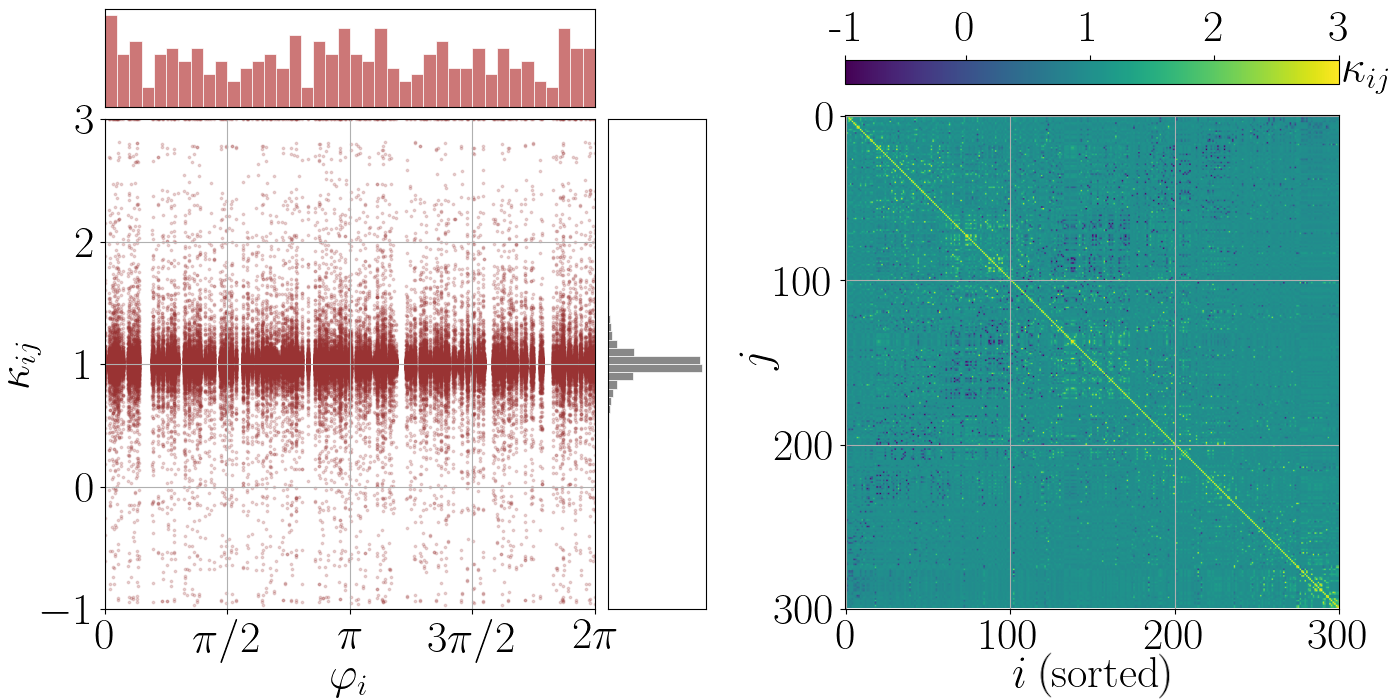"}\put(44,44){\large\bfseries $(a)$}\end{overpic}
\begin{overpic}[width=0.45\textwidth]{"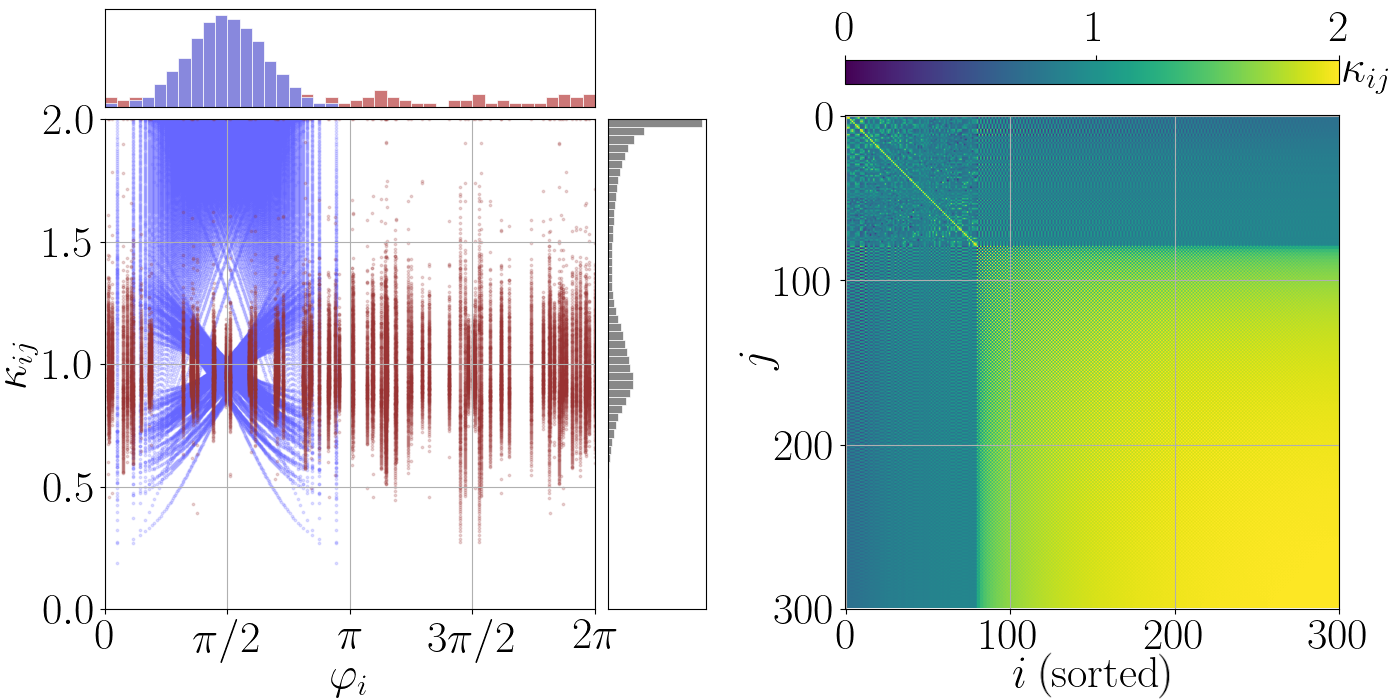"}\put(44,44){\large\bfseries $(b)$}\end{overpic}
\begin{overpic}[width=0.45\textwidth]{"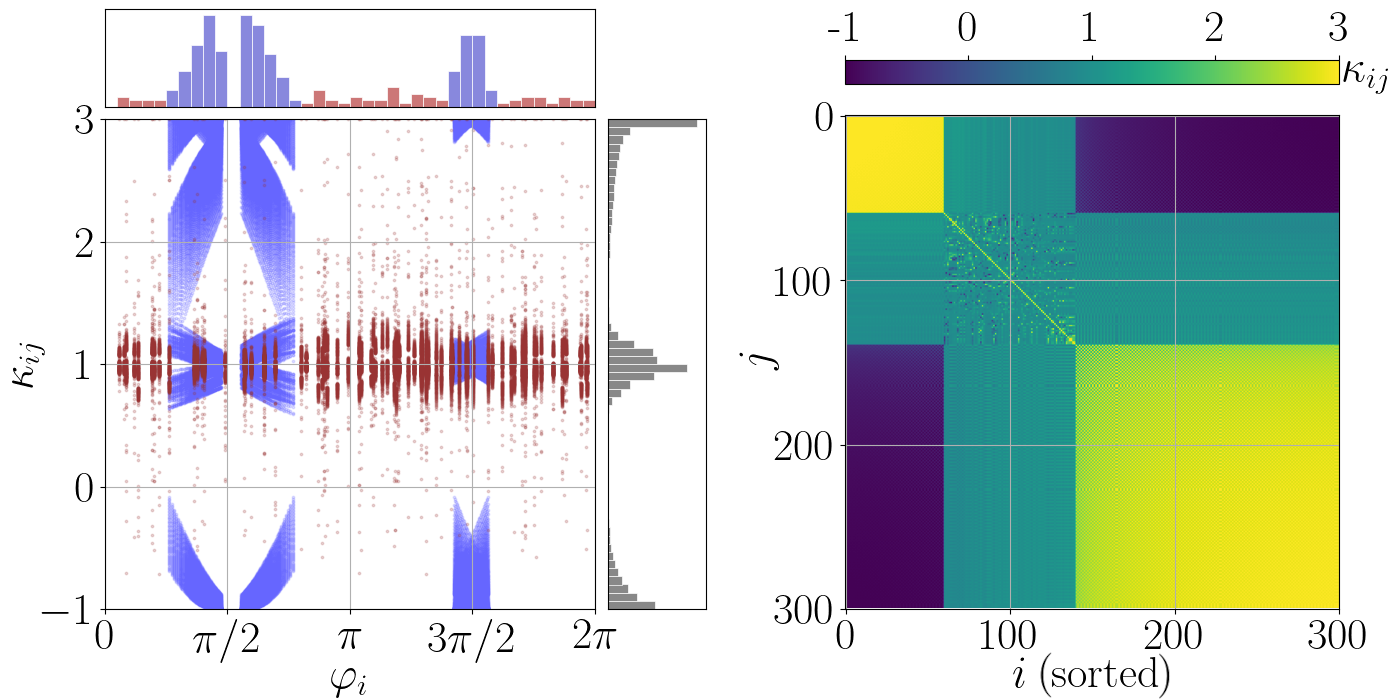"}\put(44,44){\large\bfseries $(c)$}\end{overpic}
\begin{overpic}[width=0.21\textwidth]{"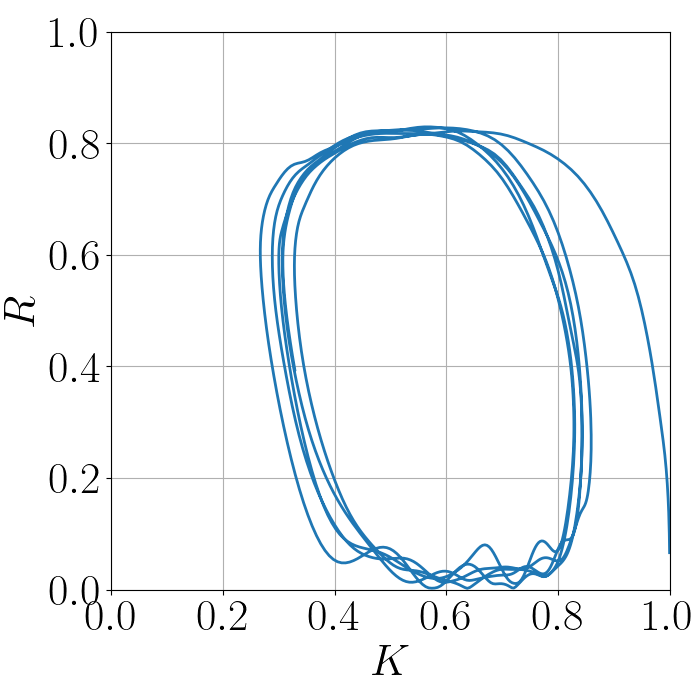"}\put(80,86){\large\bfseries $(d)$}\end{overpic}
\begin{overpic}[width=0.21\textwidth]{"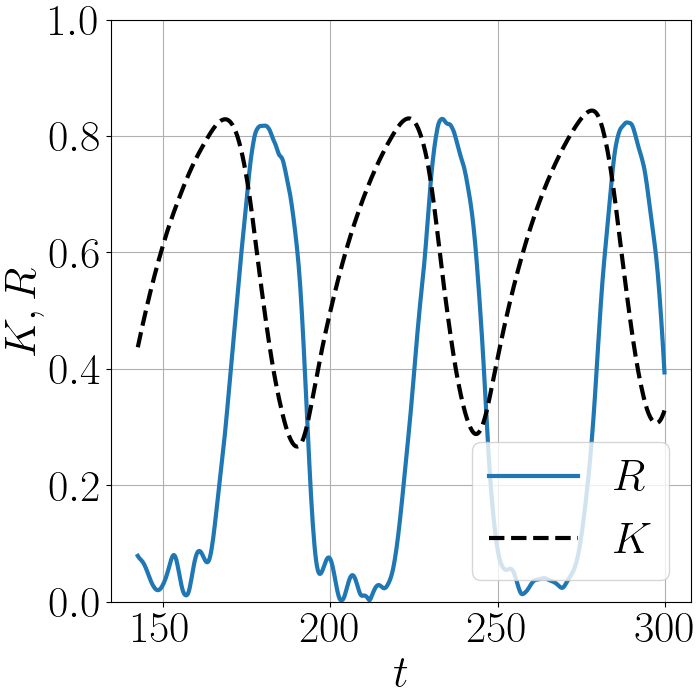"}\end{overpic}
\caption{(a-c) macroscopically stationary states observed by numerical simulation of Eqs.~\eqref{eq:full_system} using $N=300$ and $\epsilon = 0.1$. In the left panels blue are phase-locked oscillators, red a drifters. In the right panels the coupling matrices are sorted with the average edge weights to reveal the phase clusters. (d) macroscopically periodic (possibly chaotic) state. $R$ and $K$ are the order parameters of phases and coupling strengths respectively: $R = \big|\frac{1}{N}\sum_i e^{i\vp_i}\big|$, $K = \frac{1}{N^2}\sum_{i,j} \kappa_{ij}$.  Panel specific parameters are (a): $(\sigma,a) = (2, 2)$; (b): $(\sigma,a) = (1, 1)$ ;  (c): $(\sigma,a) = (1,2)$; (d): $(\sigma,a)=(0.5,-2)$. To obtain the state in panel (c), a special initial condition was selected: starting with all oscillators at the same phase, and then those with frequency close to zero were shifted by phase $\pi$, making two perfectly synced clusters (this is why we see no oscillators at the center of the main cluster at $\pi/2$, because they were used to form the other cluster). The weights within clusters were set positive, while between clusters were set negative. 
}
\label{fig:states}
\end{figure}

Through naive explorative simulations of \eqref{eq:full_system} with finite $N$ one observes three qualitatively different macroscopically-stationary states (\cref{fig:states}), all of which occur for positive adaptivity $a>0$:
\begin{enumerate}
\item{{\bf Incoherent state}: phases are scattered uniformly and most coupling weights are close to 1. This state occurs if the frequency distribution is wide enough. }
\item{{\bf One-phase cluster}: a significant portion of phases lock to the mean field and form a single coherent cluster, while the remaining phases drift and are uniformly distributed. The coupling connecting two coherent oscillators have a larger weight $\kappa_{ij}\approx 1+a$ while remaining coupling weights are close to 1. This state occurs for a narrow enough frequency distribution. }
\item{{\bf Two-phase clusters}: two coherent phase clusters form, a main cluster and a smaller cluster at the antipodal phase. The coupling  connecting oscillators of the same clusters again have large weight $\kappa_{ij}\approx 1+a$, while coupling connecting two oscillators from the opposing clusters have negative weight $\kappa_{ij} \approx 1-a$. All other coupling weights are close to 1. This state occurs for a narrow enough frequency distribution, large enough adaptivity $a$ and particular initial conditions of both phases $\vp_i$ and couplings $\kappa_{ij}$ (for incoherent initial conditions it is unlikely to reach 2-phase-cluster states).}
\end{enumerate}
See~\cref{fig:states}(a,b,c) for the depiction of these states, realized with a finite ensemble simulation $N = 300$ and $\epsilon = 0.1$. 
In the limit $N \to \infty$, provided the frequency distribution has infinite support, there is always a portion of oscillators whose phases do not lock to the mean field and so all observed quasi-stationary states feature aspects of partial synchrony~\cite{Kuramoto1975,strogatz_kuramoto_crawford_2000,martens2009exact}. However, if the frequency domain has a compact support (which is always the case in numerical realizations) one can also get fully locked states.  

For negative adaptivity $a < 0$ we also observe macroscopically non-stationary states, both periodic as well as chaotic, where the mean field changes both direction and amplitude in time, see Fig.~\ref{fig:states}(d). These states are reminiscent to those of recurrent synchronization~\cite{recurrent_synchronization} (periodic) and recurrent chaotic synchronization~\cite{chaotic_recurrent_synchronization_arxiv} (chaotic) which have recently been reported for finite sized adaptive systems. 

\section{Row-average approximation}
In system~\eqref{eq:full_system} the number of edges grows quadratically with the number of nodes which makes numerical simulations of large ensembles prohibitively slow. This is why, particularly when aiming at studying the system in the continuum limit $N\to\infty$, it is useful to consider simplifications. One such simplification was done in Ref~\cite{duchet_bick_2023} where all edge weights are averaged and represented with a single macroscopic variable. However, here we desire to pursue a less drastic reduction, while preserving the information about the internal edge dynamics of the network, and instead treat the network topology as a nodal property. Specifically, we consider the row-average of the coupling matrix,
\begin{equation}
\kappa_i = \frac{1}{N}\sum_{j=1}^N \kappa_{ij}\,.
\label{eq:kappa_i_node}
\end{equation}
Constraining the dynamics of system~\eqref{eq:full_system} onto $\kappa_i$ variables we obtain the following dynamical equations:
\begin{subequations}
\begin{align}
\dot{\vp}_i &= \omega_i + \kappa_i R \sin(\Phi-\vp_i)\,,\\
\dot{\kappa}_i &= \epsilon(1+aR\cos(\Phi-\vp_i)-\kappa_i)\,,
\end{align}
\label{eq:nodal}
\end{subequations}
where $R$ and $\Phi$ are the phase order parameters:
\begin{equation}
Re^{i\Phi} = \frac{1}{N}\sum\limits_{i=1}^N e^{i\vp_i}\,.
\end{equation}
We can move into the rotation frame of reference and without loss of generality set $\Phi = 0$:
\begin{subequations}
\begin{align}
\dot{\vp}_i &= \omega_i - \kappa_i \, R \,\sin(\vp_i)\,, \label{eq:nodal0_fi}\\
\dot{\kappa}_i &= \epsilon(1 + a \, R \, \cos(\vp_i) - \kappa_i)\,. \label{eq:nodal0_kappa}
\end{align}
\label{eq:nodal0}
\end{subequations}
These are the equations we will analyze in this paper. 

System~\eqref{eq:nodal0} is not equivalent to system~\eqref{eq:full_system}, but rather serves as its approximation. While both systems are related, \eqref{eq:nodal0} may capture only an aspect of the dynamics described by \eqref{eq:full_system}. The precise quality of this approximation is difficult to quantify analytically, and is best assessed through numerical simulations. Although we later provide basic simulations to demonstrate the relationship between the two systems, future work should explore more extensive simulations and theoretical analyses to refine this understanding. This case serves as a foundational model, similar to how the recent work of Duchet et al.~\cite{duchet_bick_2023} provides a benchmark model for the mean-field adaptive networks, offering a baseline for more complex formulations (like this paper) in future studies. 

\section{Locked and Drifting oscillators}\label{sec:locked_drifting_oscillators}
We aim at analyzing the stationary partial synchrony states observed in \cref{sec:full_system} with a self-consistency equation.
We follow Kuramoto~\cite{Kuramoto1975} and Strogatz~\cite{strogatz_kuramoto_crawford_2000} in the derivation (see Section 4 of Ref.~\cite{strogatz_kuramoto_crawford_2000}). As an overview, we will evaluate the order parameter $R$ via the integral:
\begin{equation}
R = \left| \int\limits_{-\infty}^\infty g(\omega) \int\limits_0^{2\pi} e^{i\vp} P(\vp,\omega) \d \vp\d\omega \right|\,,
\label{eq:R_general_integral}
\end{equation}
where $g(\omega)$ is the frequency distribution and $P(\vp,\omega)$ is the probability density of oscillators at phase $\vp$ and frequency $\omega$. 

\begin{figure}
\begin{overpic}[width=0.4\textwidth]{"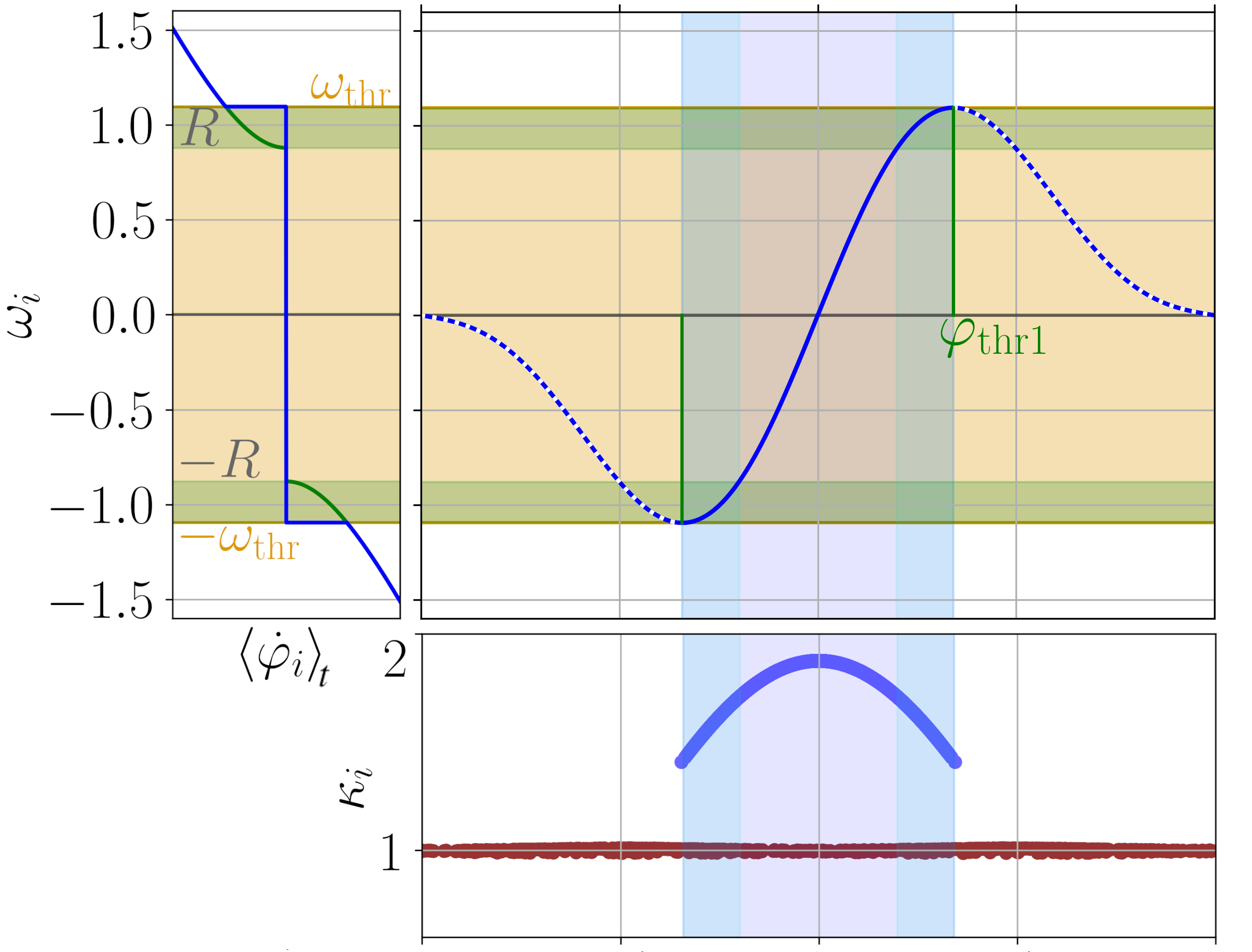"}
\put(-2,71){\large\bfseries $(a)$}
\end{overpic}
\\\smallskip
\begin{overpic}[width=0.4\textwidth]{"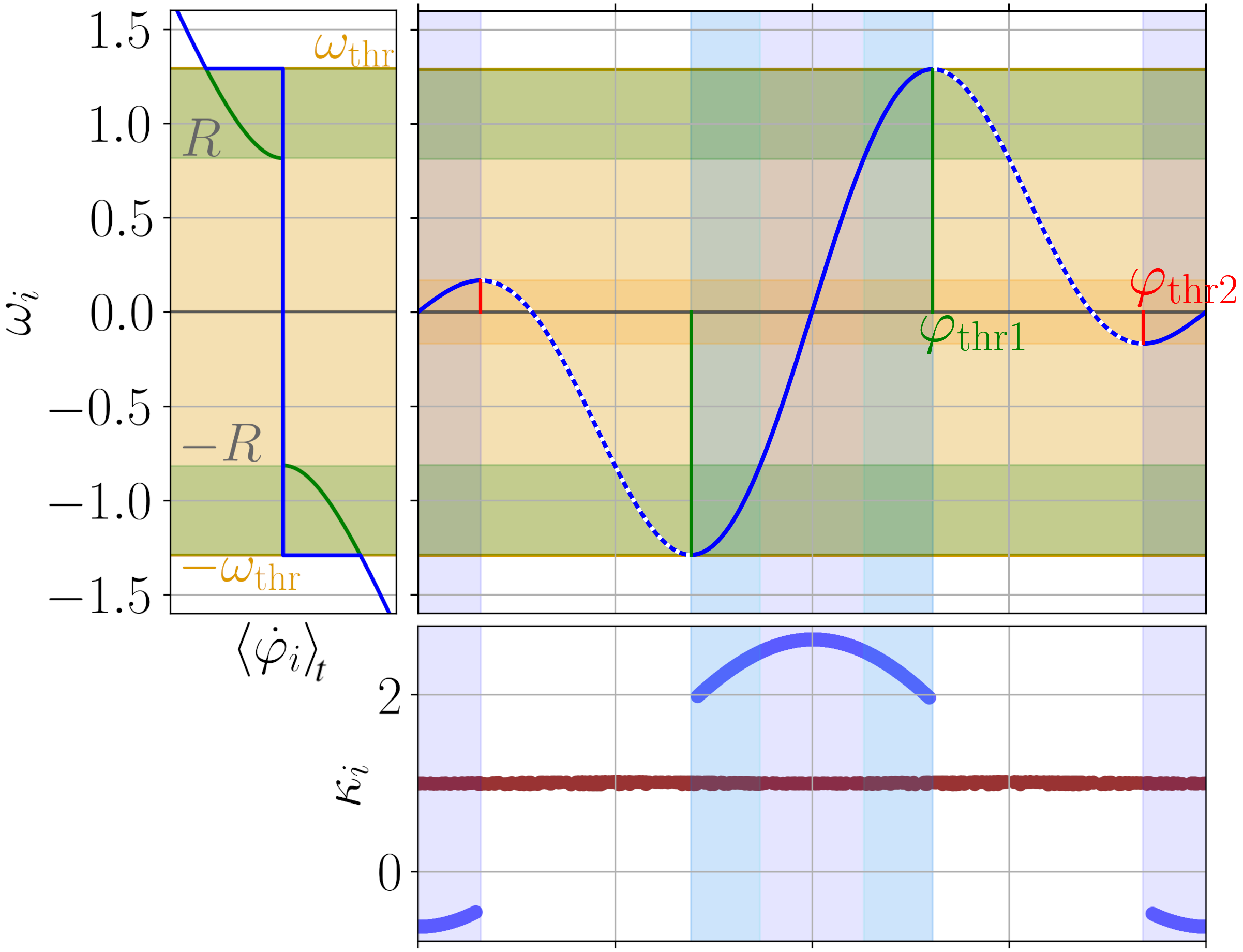"}
\put(-2,71){\large\bfseries $(b)$}
\end{overpic}
\\\smallskip
\begin{overpic}[width=0.4\textwidth]{"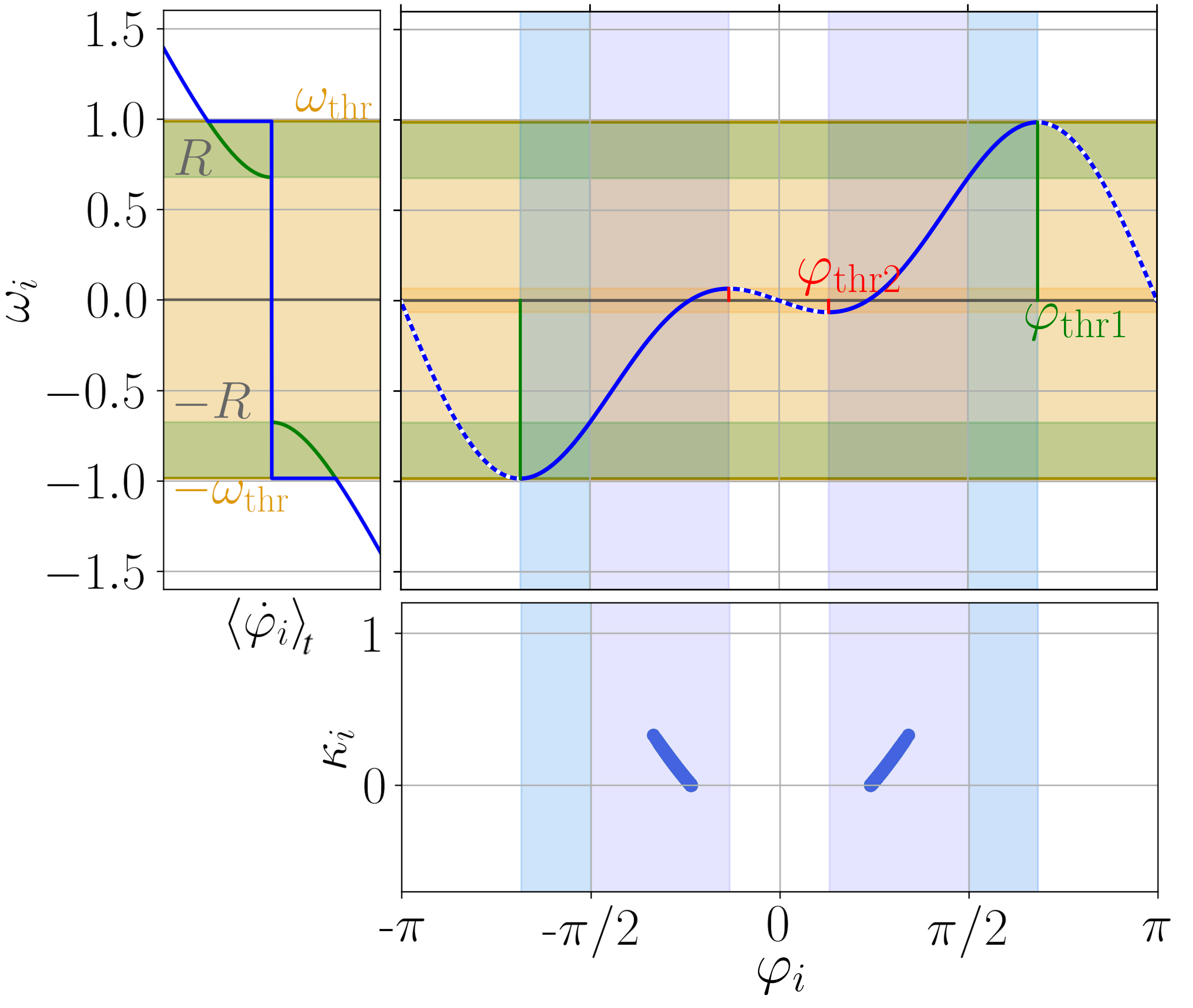"}
\put(-2,80){\large\bfseries $(c)$}
\end{overpic}
\caption{
The phase-frequency relation~\eqref{eq:w_to_fi} in three qualitatively different situations. (a) injective relation with stable phases only around 0, $(a,R,\sigma) = (1,0.88,0.6)$, (b) positive adaptation non-injectivity with additional stable phases around phase $\pi$, $(a,R,\sigma) = (2,0.81,0.65)$, (c) negative adaptation non-injectivity, splitting of stable phases on either side of phase 0, $(a,R,\sigma) = (-2,0.68,0.05)$. The $R$ values were obtained from direct numerical simulations of the states (bottom sub-panels). Additionally, each of the three panels shows the dynamic average frequency $\langle \dot{\vp}_i\rangle_t$ on the left, and an example state of system~\eqref{eq:nodal0} on the bottom with $\epsilon = 0.01$ and $N= 10^4$ (locked oscillators blue, drifters red). The shaded vertical/horizontal stripes mark particular phase/frequency domains: blue/orange marks the domain where oscillators lock and cyan/green marks the domain where oscillators can either drift or lock. The horizontal dark orange marks the frequency domain that allows locking at two distinct phases. Outside these regions oscillators drift. 
All illustrated states have configurations with maximum locking, obtained by using initial conditions with high absolute values for $\kappa_i$. The state of panel (c) has no drifters, corresponding to a fully locked state (due to finite $N$, note that the continuum limit always displays drifters). 
}
\label{fig:relation}
\end{figure}

We split the oscillator population into two groups:
\begin{enumerate}
\item{Phase-locked oscillators, denoted {\bf locked}: $\dot{\vp}_i = 0$. }
\item{Phase-drifting oscillators, denoted {\bf drifters}: $\dot{\vp}_i \neq 0$. }
\end{enumerate}
Note that for all asymptotically stable stationary states each oscillator can only either be locked or drifting. However, for $a \neq 0$ there is always a range of natural frequencies that allow oscillators to be either locked or drifting depending on initial conditions, and both oscillator states provide a multitude of locally stable configurations with varying ratio of locked versus drifting oscillators (see green region in Fig.~\ref{fig:relation}). Thus, the adaptive capability of the modified Kuramoto model in ~\eqref{eq:nodal0} allows for a high level of multistability that is absent for the non-adaptive model.
To make progress in terms of our analysis, we restrict our focus only to configurations with (i) symmetric phase oscillator distribution (i.e., $P(\vp) = P(-\vp)$ with $P$ the (quasi-stationary) probability density distribution),  and (ii) with maximal locking, i.e., the maximal amount of oscillators reside in the locked oscillator state -- in other words, we adhere to the principle that {\it "all that can lock, do lock"}. We denote the maximal possible frequency of locking with $\omega_\text{thr}$. For a discussion of challenges encountered for asymmetric distributions see  \cref{sec:asym_distr}.

We may express the order parameter $R$ as having two contributions -- due to locked oscillators and drifters:
\begin{equation}
R = R_\text{DRIFT}+R_\text{LOCK} = \Big|\int\limits_{|\omega| > \omega_\text{thr}} \!\!\!\! \cdot \ \d\omega \Big| + \Big|\int\limits_{|\omega| < \omega_\text{thr}} \!\!\!\! \cdot \ \d\omega \Big|\,.
\end{equation}
Let us first consider the {\bf drifters}, the oscillators that do not lock to the mean field and for which $\dot{\vp}_i \neq 0$. Note that we anticipate a nearly identical calculation as in the case of the classical Kuramoto system~\cite{Kuramoto1975,strogatz_kuramoto_crawford_2000}. 
Taking the limit $\epsilon \to 0$ we can average the fast phase dynamics in~\cref{eq:nodal0_kappa}~\cite{guckenheimer2013nonlinear,kuehn2015multiple}:
\begin{equation}
\dot{\kappa}_i = \epsilon(1+aR\langle \cos(\vp_i)\rangle_t -\kappa_i)\,,
\label{eq:slow_fast_averaging}
\end{equation}
where by $\langle \cdot \rangle_t$ we denote the time average limit: $\langle \cdot \rangle_t = \lim\limits_{T\to\infty} \frac{1}{T}\int\limits_0^T \cdot \d t$. 
In this averaged slow dynamics the $\kappa_i$ tend towards a certain \emph{stationary} value,
\begin{equation}
\kappa_i^\text{st} = 1+aR\langle \cos(\vp_i) \rangle_t,
\label{eq:ki_drift_stat}
\end{equation}
This is the case for the limit $\epsilon \to 0$, but for finite $\epsilon \ll 1$ this is still a good approximation - while $\kappa_i$ variables are not exactly stationary, they only weakly deviate from this value. 
We can substitute this $\kappa_i^\text{st}$ value into the phase equation~\eqref{eq:nodal0_fi} and obtain a simple ensemble of phase oscillators coupled via the first harmonic, in similarity to the analysis for the Kuramoto model with heterogeneous coupling strengths~\cite{strogatz_kuramoto_crawford_2000}:
\begin{equation}
\dot{\vp}_i = \omega_i-(1+aR\langle \cos(\vp_i) \rangle_t)R\sin(\vp_i)\,.
\label{eq:phase_eq_drift}
\end{equation}
Since we are considering drifting oscillators, we do not expect their phases to become stationary, but in the continuum limit of $ N \to \infty $ we do expect their probability density to be stationary. Their density is inversely proportional to their velocity:
\begin{equation}
P(\vp,\omega) = \frac{C}{|\dot{\vp}|} = \frac{C}{|\omega-(1+aR\langle \cos(\vp)\rangle_t) R \sin(\vp)|}\,,
\label{eq:drift_distr}
\end{equation}
(with $C \in \mathbb{R}$ the normalization constant). 
The term $\langle \cos(\vp)\rangle_t$ turns out to be zero which implies $\kappa_i^\text{st} = 1$, and so for drifters the governing phase-equations are identical to the Kuramoto model (see Appendix~\ref{sec:drifter_ki} for a derivation)~\footnote{This is why some oscillators can drift or lock depending on initial conditions since they can adjust their coupling to either $\kappa_i^\text{st} = 1$ if they are drifting or $\kappa_i^\text{st} = 1+aR\cos(\vp_i^\text{st})$ if they lock.}.

We can evaluate the integral~\eqref{eq:R_general_integral} for drifters using the expression for their probability density~\eqref{eq:drift_distr}. Importantly, keep notice of the symmetry: 
\begin{equation}
P(\vp+\pi,-\omega) = P(\vp,\omega)\,, 
\end{equation}
which will be crucial in showing that the contribution of drifting oscillators to the mean field vanishes, just like it does in the case of the standard Kuramoto model~\cite{Kuramoto1975,strogatz_kuramoto_crawford_2000}.
Additionally, remember we only consider symmetric frequency distributions $g(\omega) = g(-\omega)$. Considering the symmetries of the phase density $P(\vp,\omega)$ and frequency distribution $g(\omega)$ we observe that the integral for drifting oscillators vanishes, see Appendix~\ref{sec:drifter_R} for the derivation. Thus, drifters do not contribute to the mean field $R$ and we may ignore the $R_\text{DRIFT} = 0$ term. Note that the limit $\epsilon \to 0$ therefore gives a lower bound of $R$ for the general case, since for $\epsilon > 0$ the drifters have a positive net contribution $R_\text{DRIFT} > 0$. 

Now let us consider the {\bf locked} oscillators, for which $\dot{\vp}_i = 0$. Since their phases are  locked, their probability density is effectively a $\delta$ function in phase: $P(\vp,\omega) \propto \delta(\vp^\text{st}(\omega))$. This means the phase integral in~\eqref{eq:R_general_integral} can be evaluated at the stationary phase $\vp^\text{st}(\omega)$ and only the frequency integration remains. We will now express this stationary phase value $\vp^\text{st}(\omega)$. 

Locked phases inevitably imply that the corresponding topology variable $\kappa_i$ is also locked, $\dot{\kappa}_i = 0$. 
We can express the stationary value:
\begin{equation}\label{eq:stationary_kappa_i}
\kappa_i^\text{st} = 1+a\,R\cos(\vp_i)\,.
\end{equation} 
(Note that we could also obtain this result by evaluating the time-averaged solution~\eqref{eq:ki_drift_stat}).
Substituting~\eqref{eq:stationary_kappa_i} into the fixed point condition for the phase equations~\eqref{eq:nodal0_fi} yields the relation between phases $\vp_i$ and their corresponding frequencies $\omega_i$:
\begin{equation}
\omega = R \sin(\vp)+\frac{a\,R^2}{2}\sin(2\vp)\,.
\label{eq:w_to_fi}
\end{equation}
Note that \eqref{eq:w_to_fi} defines functional relation for each oscillator $i$, $\omega_i=f(\varphi_i)$, which is not necessarily injective, i.e., for the same frequency $\omega$ there can be more than one stable phase value $\vp$~\footnote{Expression~\eqref{eq:w_to_fi} always has 4 complex-valued solutions, but we are only interested in the real-valued ones, which can be 2 or 4. Additionally we care about stable solutions that satisfy: $\frac{\partial \omega}{\partial \vp} > 0$, since they represent viable system states. Real-valued solutions come in stable-unstable pairs so if there are 2 solutions in total then 1 is stable, and if there are 4 then 2 are stable.}. This is what gives rise to two phase cluster states, as we will see further below in the subsequent analysis. 
Since the phase-frequency relation~\eqref{eq:w_to_fi} is a second-order harmonic, we can expect at most 2-cluster states. If higher-order harmonics were introduced in either phase coupling~\eqref{eq:nodal0_fi} or the adaptation rule~\eqref{eq:nodal0_kappa} it could raise the order of the phase-frequency relation, in which case larger multi-cluster states could be expected. 
In Fig.~\ref{fig:relation} we plot relation~\eqref{eq:w_to_fi} for different adaptivity $a$. There are three qualitatively different situations:
\begin{itemize}
\item{ $|a| < 1$: \eqref{eq:w_to_fi} is injective, and the analysis is analogous to the regular Kuramoto model. Only 1-phase-cluster states are possible, see~\cref{fig:relation}(a).}

\item{ $a > 1$: \eqref{eq:w_to_fi} produces another bump around phase $\pi$, which can give rise to 2-phase-cluster antipodal states (main cluster around phase 0, and a smaller cluster around phase $\pi$), see~\cref{fig:relation}(b).
}

\item{ $a < -1$: \eqref{eq:w_to_fi} produces another bump around phase $0$ which can give rise to (potentially unstable) 2-phase-cluster states (both clusters of roughly equal size on either side of phase 0), see~\cref{fig:relation}(c).}
\end{itemize}
The threshold values that determine phase locking regions $\vp_\text{thr1},\vp_\text{thr2}$ can be expressed explicitly:
\begin{subequations}
\begin{align}
\vp_\text{thr1} &= \arccos\left( \frac{1}{4\,a\,R} \left(-1+\sqrt{1+8\,a^2\,R^2} \right) \right)\,,\label{eq:fithr1}\\
\vp_\text{thr2} &= \arccos\left( \frac{1}{4\,a\,R} \left(-1-\sqrt{1+8\,a^2\,R^2} \right) \right)\,,\label{eq:fithr2}
\end{align}
\end{subequations}
and so can the maximal locking frequency threshold:
\begin{equation}
\omega_\text{thr} = \omega(\vp_\text{thr1}) = R \sqrt{\frac{(3+\sqrt{1+8 \,a^2\,R^2})^3}{32(1+\sqrt{1+8\,a^2\,R^2})}}\,.
\end{equation}

\section{Self-consistency equation for the order parameter $R$}
We define the order parameter $R$  as the absolute value of the mean phase exponentials, i.e., Eq.~\eqref{eq:R_general_integral}. 
We have already seen that the drifters do not contribute to the order parameter and that the locked oscillators may only have one (or two) phase values for each frequency $\omega$ 
-- because we only consider maximal locking states ({\it "all that can lock, do lock"}), otherwise one could talk about having a distribution of phase values for frequencies that allow for locking and drifting (green strip in Fig.~\ref{fig:relation}). 
We can therefore express the order parameter with an integral over the frequency only:
\begin{equation}
R = \left| \,\int\limits_{-\omega_\text{thr}}^{\omega_\text{thr}} e^{i\vp(\omega)} g(\omega)\d\omega \,\right|\,.
\label{eq:R_expression_broad}
\end{equation}
In the case where there are two phase values $\vp$ for each frequency $\omega$, this expression should be interpreted more broadly, see Appendix~\ref{sec:R_integral}.
Additionally, since we only consider symmetric phase distributions (with symmetric frequency distribution $g(\omega)$) and move into the rotating frame of reference, we need only consider the real component of the order parameter (the integral over $\sin(\vp)$ cancels):
\begin{equation}
R = \int\limits_{-\omega_\text{thr}}^{\omega_\text{thr}} \cos(\vp(\omega)) g(\omega)\d\omega\,.
\end{equation}
However, it is worth noting that in general for non-symmetric phase distributions 
we have to consider the general case, see Appendix~\ref{sec:asym_distr} for details. 

We perform a change of variables using relation~\eqref{eq:w_to_fi} and re-write the integral over the phases, while keeping in mind the non-injectivity of this relation. Thus we obtain a self-consistency equation for the order parameter $R$:
\begin{align}
 \begin{split}
    R  &=  R \int\limits_{D(\vp)}  \left[\cos^2(\vp)  +  aR \cos(\vp)\cos(2\vp)  \right] \\
    &\quad \quad \times g\left(  R \sin(\vp)  +  \frac{aR^2}{2}  \sin(2\vp)  \right) \d\vp\,.
    \label{eq:scr}
 \end{split}
\end{align}
The phase domain $D(\vp)$ can be as simple as $(-\vp_\text{thr1},\vp_\text{thr1})$ but can also be non-trivial since we can have 2-phase cluster states (see blue region in Fig.~\ref{fig:relation}). Also keep in mind that a portion of oscillators can either be locked or drifting depending on their initial conditions (green strip in Fig.~\ref{fig:relation}), although we here always consider the maximal locking solution and thus {\it "all that can lock, do lock"}.

\subsection{The choice of domain $D(\vp)$}
We have to be mindful of what we mean by integrating over the phases: we know that relation~\eqref{eq:w_to_fi} is not necessarily injective, as well as some oscillators with natural frequencies $R < |\omega| < \omega_\text{thr}$ may be locked or drifting.  
We also note that we have a choice in defining the integration domain $D(\vp)$, as it corresponds to which states we are trying to capture with our self-consistency equation~\eqref{eq:scr}. For example, in the case of strong positive adaptation (Fig.~\ref{fig:relation}(b)) we can consider only one cluster states and integrate only over the main branch around $\vp = 0$,
\begin{equation}
R = \int\limits_{-\vp_\text{thr1}}^{\vp_\text{thr1}} \cdot \d\vp\,,
\end{equation}
or we can consider a portion of oscillators to lock at a different phase and thus contribute to the second (antipodal) phase cluster. Only oscillators within a particular range of frequencies have this choice (see thin dark orange stripe in Fig.~\ref{fig:relation}(b)). Specifically, suppose oscillators with frequencies ranging from $-\omega^\dagger$ to $\omega^\dagger$ go in the antipodal cluster. We can choose $\omega^\dagger$ as the parameter. To write the integral in terms of the phase we then have to find the corresponding phase threshold values which we compute numerically by satisfying Eq.~\eqref{eq:w_to_fi} with $\omega_\dagger$ on the left-hand side. 
This yields two phase solutions, near zero: $\vp_0^\dagger$ and near $\pi$: $\vp_{\pi}^\dagger$. 
The integral is then written as follows:
\begin{equation}
R = \int\limits_{-\vp_\text{thr1}}^{\vp_\text{thr1}} \cdot \d\vp \ \ + \int\limits_{\vp_{\pi}^\dagger}^{2\pi-\vp_{\pi}^\dagger} \cdot \d\vp \ \  - \int\limits_{-\vp_0^\dagger}^{\vp_0^\dagger} \cdot \d\vp\,.
\label{eq:int_antipodal2}
\end{equation}

For strongly negative adaptation we always get two phase-clusters (see Fig.~\ref{fig:relation}(c) but we still have to make a choice of the integration domain. We here only consider the symmetric solutions and an injective relation~\eqref{eq:w_to_fi} where all negative frequencies go to one cluster and all positive to the other (this appears to be the common attractor for starting with a completely incoherent state). We have to look for the phase value that is the non-trivial zero of relation~\eqref{eq:w_to_fi}: $\vp^*$, and the self-consistency integral is now written as:
\begin{equation}
R = \int\limits_{-\vp_\text{thr1}}^{-\vp^*} \cdot \d\vp \ \ + \int\limits_{\vp^*}^{\vp_\text{thr1}} \cdot \d\vp\,.
\end{equation}

\section{Phase diagram}
We analyze the row-averaged system~\eqref{eq:nodal0} based on solving the self-consistency equation~\eqref{eq:scr} using analytical (\cref{sec:transition_to_sync}) and numerical methods (\cref{sec:iterating_scr}) to determine the associated stability diagram shown in Fig.~\ref{fig:boundaries}. To examine non-stationary phase configurations, these results are further complemented by numerical simulations of \cref{eq:nodal0} in \cref{fig:boundaries}(a). 

\subsection{Incoherence-coherence transition}\label{sec:transition_to_sync}
\begin{figure}
\begin{overpic}[width=0.38\textwidth]{"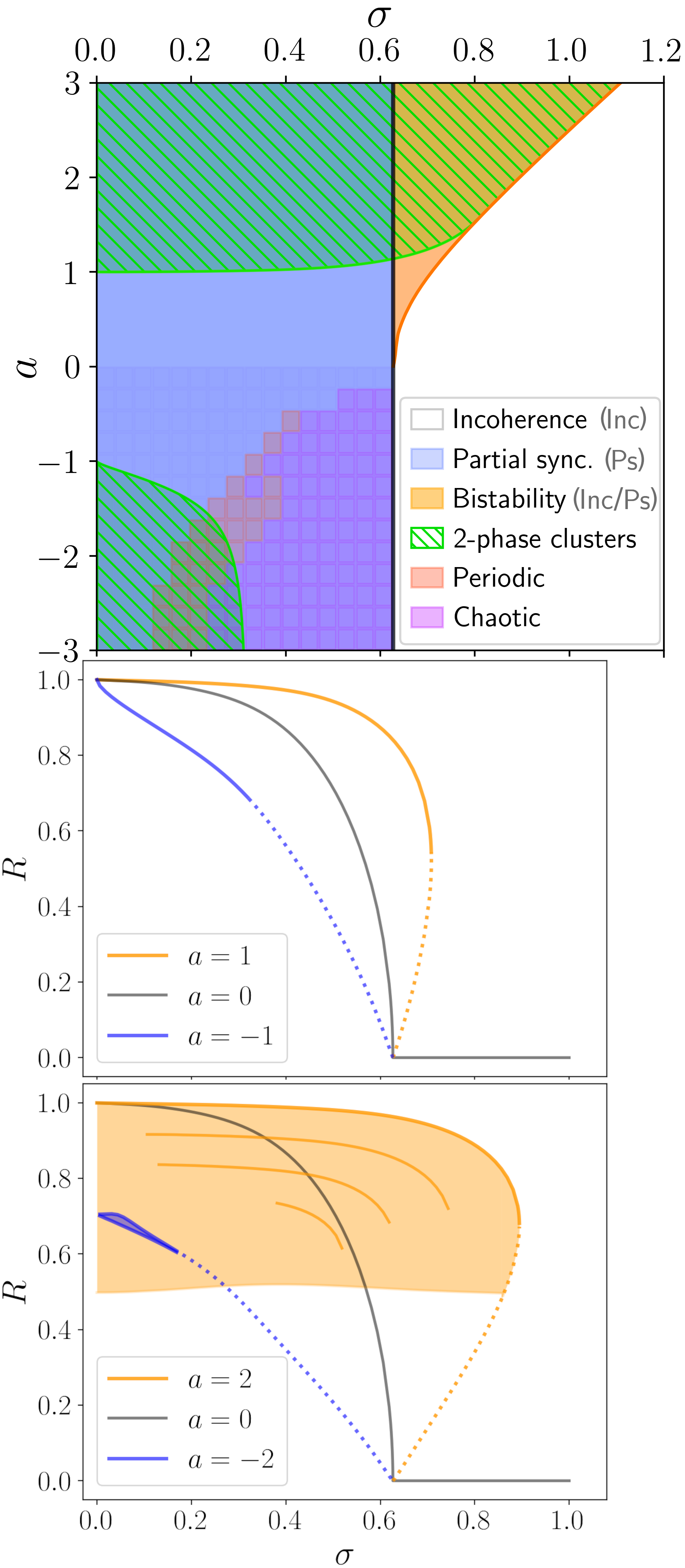"}
    \put(-2,91){\large (a)}
    \put(-2,54){\large (b)}
    \put(-2,27){\large (c)}
    \put(36,89){\rotatebox{45}{F}}
    \put(28.4,45){F}
    \put(26.6,45){ \tikz \draw[orange,fill=orange] (0,0) circle (.4ex);} 
    \put(34,21){F}
    \put(32.2,21.2){ \tikz \draw[orange,fill=orange] (0,0) circle (.4ex);} 
    \put(22.2,92.8){ $\sigma_c$}
    \put(20,19){\textcolor{orange}{$c_3$}}
    \put(20.8,20){ \tikz \draw[orange,fill=orange] (0,0) circle (.4ex);} 
    \put(22,22){\textcolor{orange}{$c_2$}}
    \put(23.9,21.5){ \tikz \draw[orange,fill=orange] (0,0) circle (.4ex);} 
    \put(24,24.7){\textcolor{orange}{$c_1$}}
    \put(27.7,22.){ \tikz \draw[orange,fill=orange] (0,0) circle (.4ex);} 
    \put(26,27){\textcolor{orange}{$c_0$}}
    \put(10.44,19.7){ \tikz \draw[blue,fill=blue] (0,0) circle (.4ex);} 
    \put(14.9,48.7){ \tikz \draw[blue,fill=blue] (0,0) circle (.4ex);} 
\end{overpic}
\caption{
(a) Phase diagram. The transition from the incoherent to the partial synchronous state at $\sigma_c = \sqrt{\pi/8}$ (black line) is independent of adaptivity $a$. 
For positive adaptivity $a>0$ a fold bifurcation (orange curve F) is seen in the $\sigma > \sigma_c$ region. A region of bi-stability (orange) between incoherence and partial synchrony lies between the fold bifurcation and transition at $\sigma_c$. 
2-phase-cluster states with high level of multistability can occur in a parameter region (green hatching) characterized by a non-injective phase frequency relationship~\eqref{eq:w_to_fi}.
For $a>0$ all states are stationary; more intricate dynamics are possible for $a<0$: numerical integration of Eqs.~\eqref{eq:nodal0} reveals stationary solutions (blue squares) for small $\sigma$; periodic solutions (red) for larger $\sigma$; and chaotic dynamics for even larger $\sigma$ (purple). 
(b),(c) Order parameter $R$ as function of $\sigma$ by solving the self-consistency equation~\eqref{eq:scr}, 
and the classical Kuramoto transition (black) as comparison. 
(b) Intermediate adaptivity: Positive $a=1$ (orange curve) reveals a fold bifurcation (F), which is absent for negative adaptivity (blue curve for $a=-1$). 
(c) Stronger adaptivity:  For positive $a=2$ we observe a region of dense multistability: neutrally stable branches (orange curves) corresponding to non-trivial 2-phase-clusters  with varying $c$ values (see \cref{sec:pos_adapt_c}) are shown: 
$c_0=0$, $c_1 = 0.05$, $c_2 = 0.1$, $c_3 = 0.3$. 
For negative $a=-2$ the stationary branch is mostly unstable (dotted blue curve) with periodic/chaotic solutions. For small $\sigma$ a narrow region with 2-phase-clusters is observed. All depicted branches represent maximal locking configurations, the range of different $R$ values is due to different proportions of oscillators in the antipodal cluster.
}
\label{fig:boundaries}
\end{figure}
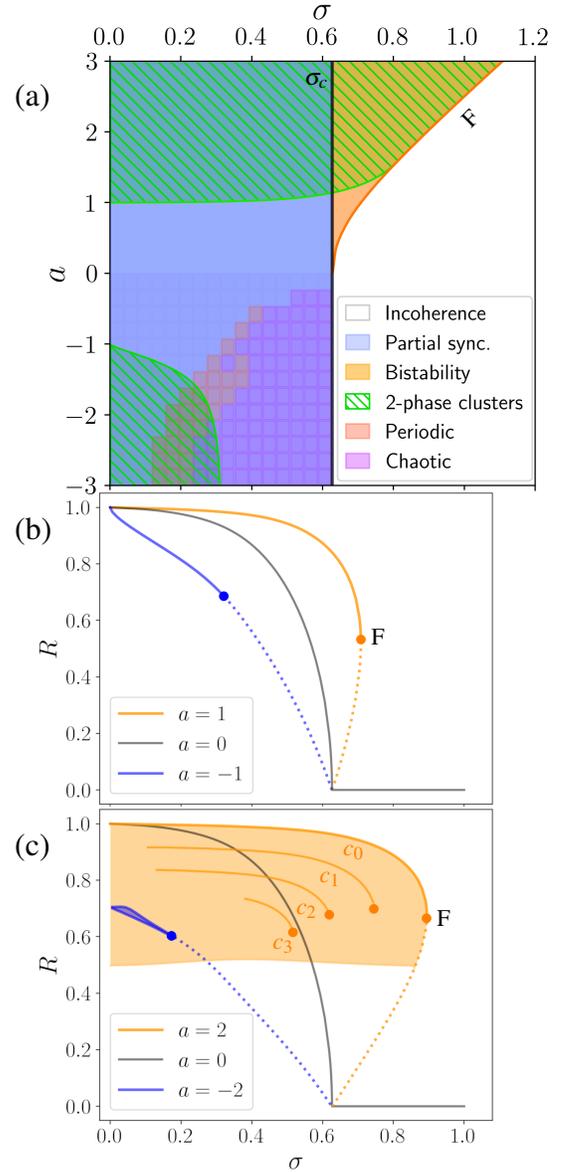

Looking at the self-consistency equation~\eqref{eq:scr} we can see that, irrespective of parameter choice,  $R=0$ is always a solution, corresponding to incoherent oscillations. The questions are whether this solution branch for the self-consistency equation is is the only one or if other solution branches exist, and if the branch $R=0$ describes a stable oscillator configuration.
Following the steps of Kuramoto's original analysis~\cite{Kuramoto1975,strogatz_kuramoto_crawford_2000}, we consider~\eqref{eq:scr}  in the limit $R\to0$ which yields
\begin{equation}\label{eq:SC_reduced}
1 = g(0)\int\limits_{D(\vp)} \cos^2(\vp) \d\vp\,,
\end{equation}
for which the integration domain in this case is $(-\pi/2,\pi/2)$. Studying this reduced self-consistency equation, we may find a condition for which another solution arises. The integral in \eqref{eq:SC_reduced} evaluates to $\pi/2$ so that the transition only depends on the shape of the frequency distribution, more specifically, the value of the frequency distribution at $\omega = 0$, 
\begin{equation}
g(0) = \frac{2}{\pi}
\end{equation}
Remember that we consider the normal distribution for $g(\omega)$ and as such $g(0)$ is just determined by the width of the distribution $\sigma$: $g(0) = \frac{1}{\sigma\sqrt{2\pi}}$. 
The critical value for the onset of a non-trivial solution branch of $R$ is therefore at
\begin{equation}\label{eq:sigma_crit}
\sigma_c = \sqrt{\frac{\pi}{8}} \approx 0.6266 \ldots
\end{equation}
We indicate this critical $\sigma_c$ as a black vertical line in Fig.~\ref{fig:boundaries}(a).
Furthermore, we ask what is the shape for small $R$ near this transition point. Expanding \eqref{eq:SC_reduced} in terms of $\sigma$,  we find that 
\begin{equation}\label{eq:SC_R_expansion}
R = \frac{1}{a} \frac{\sigma-\sigma_c}{C} + \mathcal{O}(\sigma^2)\,,
\end{equation}
where $\sigma_c = \sqrt{\frac{\pi}{8}}$ and $C = \int \cos(\varphi)\cos(2\varphi)d\varphi/\sqrt{2\pi} = \sqrt{\frac{2}{9\pi}}$.

\subsection{Non-adaptive limit ($a=0$) and small adaptation ($a\approx 0$)}
We continue by explaining the dynamics for the case of no adaptation, which is identical to the dynamics observed for the classical Kuramoto model. There are only two regions: the incoherent region where $R = 0$; and a partially synchronous region where $R > 0$. The transition from incoherence to partial synchrony was analytically determined by~\cref{eq:sigma_crit}, shown as a black line in Fig.~\ref{fig:boundaries} (b,c). The shape of the order parameter (gray curve in \cref{fig:boundaries}(b) and (c), \cref{fig:numerics}(a) and (b)) near this transition is approximated by~\cref{eq:SC_R_expansion}.

Any non-zero amount of adaptation  allows some oscillators to reside in either locked or drifting configurations, consistent with a certain range of natural frequencies (see \cref{sec:locked_drifting_oscillators}). Thus for non-zero adaptivity, $a\neq 0$, there always is a spectrum of neutrally stable configurations for the oscillators to occupy, corresponding to varying ratios of locked versus drifting oscillators that result in slightly different $R$ values~\footnote{The states are stable against small perturbations of any oscillator, but in the continuum limit any perturbation to a single oscillator can be considered small to the system so we call them neutrally stable.}. As mentioned previously (\cref{sec:locked_drifting_oscillators}), we always focus on the maximal locking configurations (Fig.~\ref{fig:boundaries} depicts only branches with maximal locking).

\subsection{Positive adaptation: bistability of partial synchrony/incoherence and 2-phase-cluster states}\label{sec:pos_adapt_c}
With positive adaptation, the domain of partial synchrony extends to the region with $\sigma>\sigma_c$ until it disappears in a fold bifurcation (orange curve in Fig.~\ref{fig:boundaries}(a)). In this region, configurations of partial synchrony and incoherence co-exist as stable states (orange region in Fig.~\ref{fig:boundaries}(a)). 
When adaptivity $a$ is raised above a critical value, we observe additional 2-phase cluster states (hatched green region in Fig.~\ref{fig:boundaries}(a)). For each parameter choice where 2-phase clusters occur, they form a continuum of coexisting 2-phase-cluster states that occupy a range of different $R$ values (shaded orange region in Fig.~\ref{fig:boundaries}(c)). The full lines within the orange shaded regions in Fig.~\ref{fig:boundaries}(c) represent specific choices for 2-phase-cluster state branches, obtained by solving the self-consistency equation~\eqref{eq:scr}. The hatched boundary in Fig.~\ref{fig:boundaries}(a) was found by expressing the self-consistency equation in terms of $a$ for the condition $R = \frac{1}{a}$ and solving it, see Appendix~\ref{sec:iterating_scr}. Note that all states are macroscopically stationary partial synchrony states. 

Due to a high degree of multistability, different state branches are not uniquely determined by $(\sigma,R)$ since the configuration of oscillators within the two clusters plays a role as well. We chose particular configurations of oscillators where all oscillators within a frequency range $(-\omega^\dagger,\omega^\dagger)$ occupy the antipodal cluster (cf. integral ~\eqref{eq:int_antipodal2}). When considering a branch with respect to changing parameter $\sigma$, the range $\omega^\dagger$ should scale linearly: $\omega^\dagger = c \sigma$, and the parameter $c \geq 0$ therefore parametrizes different branches. $c=0$ means that none of the oscillators are in the antipodal cluster, and therefore $c=0$ parametrizes the one-cluster state. $c$ also cannot be arbitrarily large since $\omega^\dagger$ cannot be arbitrarily large for each $\sigma$.  

\subsection{Negative adaptation: 2-phase-cluster states and non-stationary states}
With negative adaptation we do not observe any bistability regions between partially synchronous  and incoherent states, since the asymptotic behavior of $R$ around the critical point $\sigma_c$ has a negative slope (see blue curve in Fig.~\ref{fig:boundaries}(b)). 
At every negative $a$ value there seems to be a range of $\sigma$ that produce non-stationary oscillator states. To characterize these states we sampled parameter space on a grid by numerically simulating \cref{eq:nodal0} with $N=10^4$ oscillator nodes. Thus, it was possible to determine whether or not observed states correspond to stationary, periodic or chaotic oscillator configurations (blue, red and purple grid cells in Fig.~\ref{fig:boundaries}(a)). Furthermore, for strong negative adaptation, one may also observe 2-phase-cluster states (hatched green region in Fig.~\ref{fig:boundaries}(a)); however, these state appear only to be stationary for a rather narrow frequency distributions. For these 2-phase-cluster states we also see a family of solution branches featuring a range of different $R$ values (blue shaded region in Fig.~\ref{fig:boundaries}(c)), but this variation is much less pronounced when compared to positive adaptivity $a$ (orange shading). 

\section{Validation of self-consistency with numerical simulations}\label{sec:numsim}
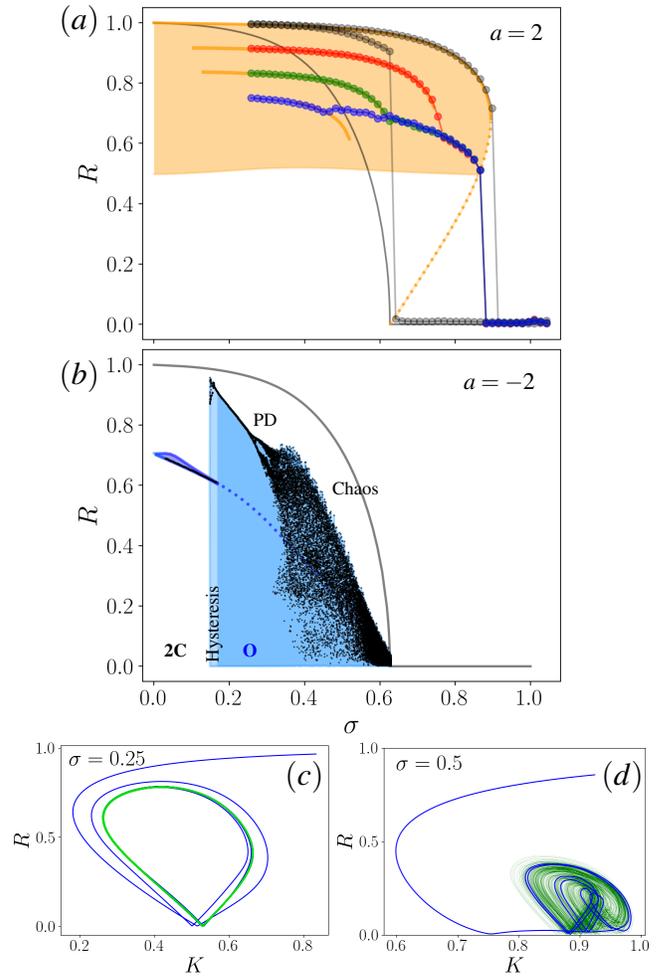
\begin{figure}
\begin{overpic}[width=0.39\textwidth]{"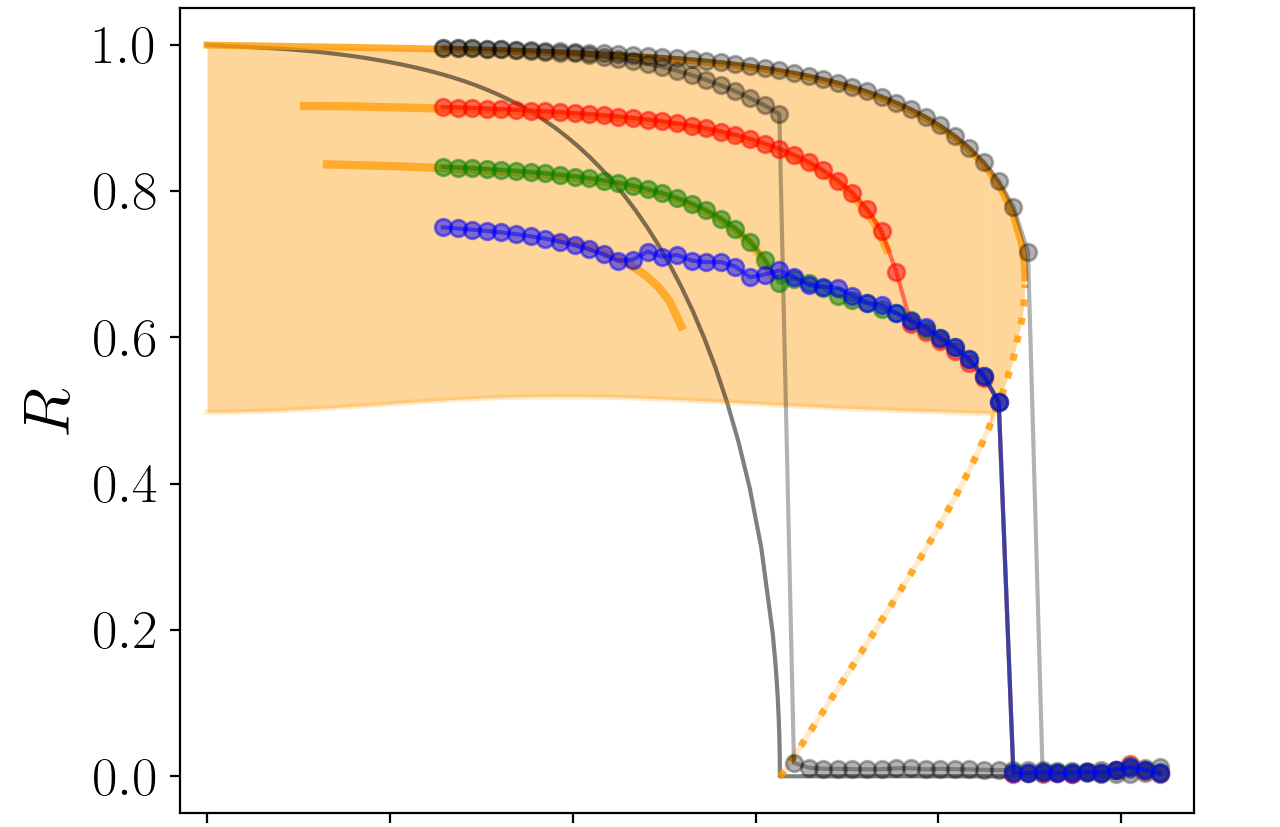"}
\put(-2,60){\large\bfseries $(a)$}
\put(80,58){$a=2$}
\end{overpic}
\begin{overpic}[width=0.39\textwidth]{"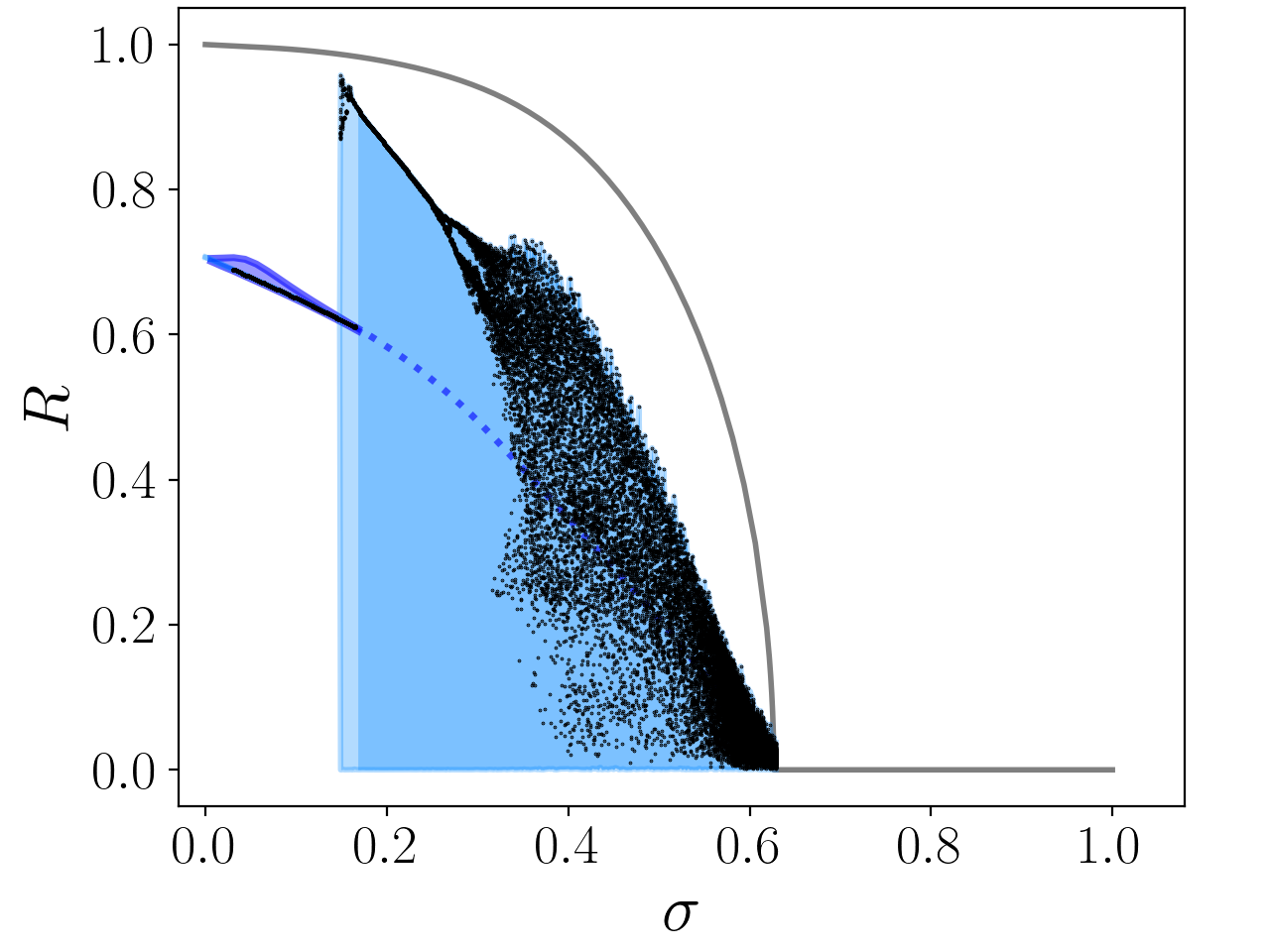"}
\put(-2,68){\large\bfseries $(b)$}
\put(75,67){$a=-2$}
\put(26,15){\scriptsize  \rotatebox{90}{Hysteresis}}
\put(18,16){\scriptsize  \bf 2C}
\put(33,16){\scriptsize  \bf \textcolor{blue}{O}}
\put(35,60){\scriptsize  PD}
\put(50,47){\scriptsize  Chaos}
\end{overpic}
\begin{overpic}[width=0.235\textwidth]{"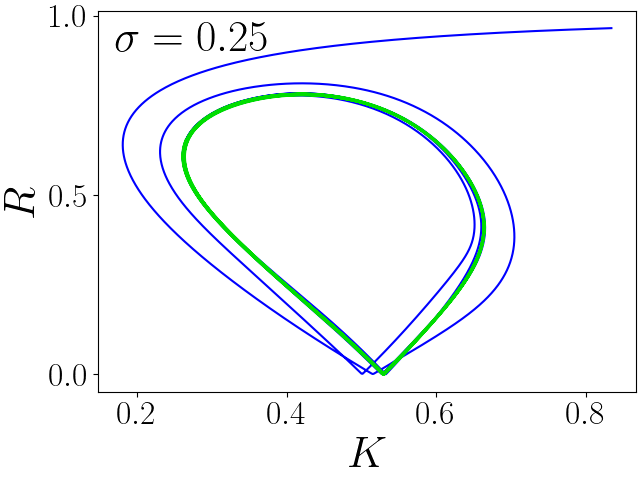"}\put(86,61){\large\bfseries $(c)$}\end{overpic}
\begin{overpic}[width=0.235\textwidth]{"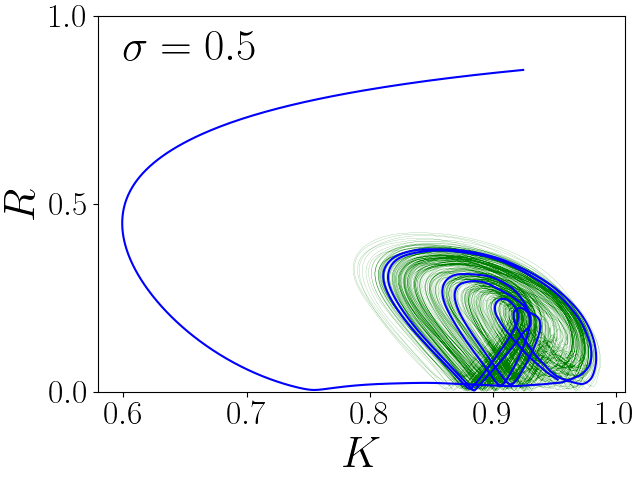"}\put(84,61){\large\bfseries $(d)$}\end{overpic}
\caption{(a),(b) Bifurcation diagrams obtained via quasi-continuation of the row-averaged model~\eqref{eq:nodal0} with $N=10^4$ oscillators and $\epsilon = 0.1$. 
(a) For positive adaptation, $a = 2$: several sweeps using different initial conditions (see text) reveal a family of neutrally stable solution branches (black, red, green, blue circles), matching well the solutions of the self-consistency equation with varying values of $c$ (orange curves, same values as in ~\cref{fig:boundaries}(c)). Quasi-continuation in forward/backward direction reveals hysteresis (black circles). 
(b) Negative adaptation, $a = -2$: forward and backward continuation reporting $\max_t{R}$  reveal stable stationary solutions (2C), and periodic solutions (O, blue shaded region) undergoing a cascade of period doubling bifurcations (PD) leading to chaotic dynamics (Chaos). A narrow region around $\sigma = 0.16$ with  bistability between stationary and non-stationary solutions is observed (Hysteresis). 
(c) Example trajectory of a periodic state with $a = -2$.
(d) Example trajectory of a chaotic state with $a = -2$. 
}
\label{fig:numerics}
\end{figure}

We compare our results obtained with the self-consistency equation~\eqref{eq:scr} with direct simulations of the finite row-averaged system~\eqref{eq:nodal0}, using $N=10^4$ oscillators with a time scale separation of $\epsilon = 0.1$. To do this, we numerically integrated \cref{eq:nodal0} and performed a quasi-adiabatic continuation while varying the parameter $\sigma$. The frequencies were deterministically sampled from a Gaussian distribution with unit variance, and then re-scaled with the parameter $\sigma$ while performing the continuation. Results are shown in~\cref{fig:numerics}.

For positive adaptation, we consider $a = 2$ and perform forward and backward quasi-continuation sweeps in $\sigma$ (black circles in \cref{fig:numerics}(a), starting from the incoherent state at $\sigma = 1.1$ and decreasing $\sigma$. We observe that the transition from incoherence to partial synchrony occurs roughly at the critical value $\sigma_c$; the order parameter jumps to a high value, but not to the maximally possible value -- this is because it goes to a state of less that maximal locking, i.e., some oscillators that could lock remain drifting.  As we continue to decrease $\sigma$ the branch merges with the branch of maximal locking (top orange curve). 
Now we reverse the direction of $\sigma$ continuation and see that the system follows the top stable branch of maximal locking and drops  to incoherence at the fold bifurcation (F) as expected (see black points in Fig.~\ref{fig:numerics}(a)) -- i.e., we observe a clear hysteresis loop. 

We also performed three other sweeps (red, green and blue circles in \cref{fig:numerics}(a)) starting from a narrow distribution at $\sigma = 0.25$ with a 2-phase cluster initial condition with different sizes of the antipodal cluster~\footnote{For initial conditions $\kappa_i$ are set to a high positive value if $|\omega_i| < \omega_\text{thr}$ and to a high negative value otherwise. The $\omega_\text{thr} = c\sigma$ determines the size of the antipodal cluster.}. The branches from numerical simulation  follow the branches obtained from the self-consistency equation for the most part; but then suddenly collapse onto a branch they all seem to share in common. Finally, this branch ceases to exist in a fold bifurcation and the incoherent state is obtained.  

For negative adaptation, we consider $a = -2$ and perform a sweep of the partially synchronous region in both directions, see Fig.~\ref{fig:numerics}(b) (starting at the critical $\sigma_c$ with incoherent initial conditions). The region between the minimum and maximum value of the order parameter $R$ is shaded blue. Stationary solutions occurring for small $\sigma$ values are distinguished from the non-stationary solutions seen for larger $\sigma$ values. We also plot the time-local maxima of the order parameter, $\max_t{(R)}$, to be able to distinguish periodic and chaotic solutions. Additionally, we notice a small hysteresis around $\sigma \approx 0.16$ where there seems to be a co-existence of the stationary and non-stationary solutions. We expect that this bifurcation diagram depends on the specific choice of parameter $\epsilon$. 

\section{Discussion}
In this paper, we investigated the dynamics of the adaptive Kuramoto model in the continuum limit with slow adaptation. Key highlights include the identification of different states, including two-cluster states within dense regions of multistability, and the finding that some oscillators can either lock or drift depending on initial conditions. We introduced a reduced approximative model, simplifying the analysis for large systems. We derived a self-consistency equation for the order parameter, revealing the conditions for the transition to synchrony. We presented a stability diagram, highlighting the effects of positive and negative adaptivity, including regions of bistability and multistability for positive adaptivity, and non-stationary states for negative adaptivity. Finally, we validated our theoretical findings through numerical simulations, confirming the predicted behaviors and transitions.

The adaptive Kuramoto model we studied introduces significant differences compared to its classical non-adaptive counterpart. While the classical Kuramoto model produces only asymptotically stationary states (on the macroscopic scale) of either complete incoherence or partial synchrony, the adaptive model exhibits a richer variety of behaviors, including macroscopically non-stationary states (see Fig.~\ref{fig:states}(d) and Fig.~\ref{fig:numerics}(c,d)). Additionally the adaptive model shows dense multistability of states for the same parameters but different initial conditions. Notably, it can produce clear two-phase cluster states (Fig.~\ref{fig:states}(c), Fig.~\ref{fig:relation}(b)) and generally asymmetric phase distributions, unlike the symmetric unimodal phase distributions observed in the classical model (for symmetric frequency distributions). This highlights the enhanced complexity and dynamic richness of the adaptive Kuramoto model. 
Moreover, the high degree of multistability in the adaptive Kuramoto model suggests potential applications in memory tasks, as the system's ability to exist in numerous states can be harnessed to represent multiple memory configurations, thereby facilitating learning and information storage.

On the microscopic scale, the adaptive Kuramoto model also shows significant differences from the classical model. In the Kuramoto model, each phase can have either only one fixed point or no fixed points, which determines whether it locks or drifts respectively. However, in the adaptive model, there is a range of natural frequencies that allow oscillators to either drift or lock depending on their initial conditions (see Fig.~\ref{fig:relation}). This ability to either drift or lock contributes to the multistability of states observed on the macroscopic scale. 

Comparing our work to Seliger et al.~\cite{adaptive_example1}, they use an adaptation rule without a constant term in Eq.~\eqref{eq:full_system_k}, which implies their model does not generalize the classical Kuramoto model if adaptation vanishes (such models have since been considered in several works~\cite{chaotic_recurrent_synchronization_arxiv,recurrent_synchronization,10.1063/1.5097835}). Without this term, Seliger's rule yields a pure second-harmonic phase coupling for stationary solutions, while our rule yields mixed first- and second-harmonic coupling. Additionally, Seliger et al. study a finite ensemble with uniformly distributed frequencies, allowing for fully stationary states. In contrast, we consider the thermodynamic limit $N\to\infty$ with normally distributed frequencies, which always result in partial synchrony and aligns our work more closely with Kuramoto's study~\cite{Kuramoto1975,strogatz_kuramoto_crawford_2000}. 

We now compare our reduced row-averaged model~\eqref{eq:nodal0} to the full $N^2$ edge model~\eqref{eq:full_system}. 
All discussed phenomena (non-stationarity, dense multistability and 2-phase cluster states) are present in both models. 
While it is possible that the full model exhibits additional phenomena not captured by the reduced version, we did not observe any and expect that all the interesting features one might find in the full model also appear in the reduction in some form. 
There are certainly quantitative differences between the two however. The two models are most similar under conditions of small adaptivity $a$, near incoherence ($R\approx0$), or near high synchrony ($R\approx1$) single cluster states. 
Settings where they differ most involve high adaptivity and pronounced 2-phase cluster states. For example, here we describe the most difference-contrasting setting we can think of: $a\gg1$ and a nearly 50-50 split 2-phase cluster state. In the full model, the main cluster at phase zero would have very high $\kappa_i$ values, and the other cluster at phase $\pi$ would have highly negative $\kappa_i$ values. This state is very stable in the full model, with the first order parameter $R$ close to zero and the second order parameter $R_2 = \langle e^{i2\vp} \rangle$ close to one. In the reduced model, such a state would not be as pronounced and might not be able to sustain such an even split due to $R\approx 0$.

The high dimensionality inherent to adaptive networks raises the question whether we can study reduced models instead, while appropriately capturing the complexity of the dynamics. 
It is therefore interesting to compare different methods of dimensional reduction. Our reduced row-averaged model~\eqref{eq:nodal0} contrasts with the even more reduced model by Duchet {\it et al.}~\cite{duchet_bick_2023}~\eqref{eq:Duchet} in several ways.
The reduction used in the work by Duchet {\it et al.} relies on averaging the coupling matrix to a single dynamic variable and combining this with Ott-Antonsen reduction applied to the phase dynamics~\cite{OttAntonsen,bick2020}. Thus, both the coupling and phase dynamics are each reduced to single mean field variables. 
As such, the resulting approximated two dimensional dynamics of the single population Duchet model (\cref{sec:Duchet}) cannot exhibit chaos; however, the model can exhibit macroscopically periodic states (see Appendix~\ref{sec:Duchet}). The macroscopic model only represents single cluster states and cannot display any microscopic effects such as oscillators having a choice of locking or drifting, nor can it represent the associated coupling strengths.
Moreover, the Duchet model does not  display non-stationary states for $a>-1$ (\cref{sec:Duchet}); this stands in contrast to our larger dimensional model relying on row-averaging. This observation illustrates well the compromises one may pay for any type of approximate dimensional reduction.
However, due to its simplicity one can consider a multi-population Duchet model, which assumes there are several populations of oscillators. The dynamics of each population is represented by its own order parameter, which interact via the multi-population coupling. 
To represent simple 2-phase-cluster states at least a two population model is required. The downside is that one has to choose the proportions of locked versus drifting oscillators in each population as a predetermined parameter (similar to choosing the value of $c$ in \eqref{eq:scr}); 
By contrast, our row-averaged model settles into one of a family of phase (and coupling) configurations depending on initial conditions; an example are 2-phase-cluster states, where individual oscillators reside in one of two specific phase configurations (while at the same time, they can reside in drift versus locked phases, too). Thus, this proportion has the character of an emergent property. 

We now compare the continuum limit to the previous work of only two adaptive oscillators~\cite{juttner_martens_2023}, highlighting the key similarities and differences. For $N=2$, Eqs.~\eqref{eq:full_system} reduce to
\begin{align*}
\dot{\vp}&=\omega - K\sin(\vp),\\
\dot{\kappa}&=\epsilon(1+a\cos(\vp)-\kappa),\
\end{align*}
where $\vp=\vp_1-\vp_2$ and $\omega=\omega_1-\omega_2$ are the phase difference and frequency mismatch of the oscillators. 
The adaptivity parameter $a$ has the same role in both models, but the width of the frequency distribution $\sigma$ does not have a one-to-one analog in the case of two oscillators, but we can relate it to the frequency mismatch of the two oscillators. We focus on figures Fig.~3(a) and Fig.~3 in Ref.~\cite{juttner_martens_2023} as they offer the clearest comparison. We identify three points of comparison: (I) for positive adaptation, the domain of synchronization increases with adaptation in both models. This can easily be rationalized for $N=2$ oscillators, where the condition for locking is $|\omega|\leq|\kappa^\text{st}|$ where $\kappa^\text{st} = 1 + a \cos(\vp^\text{st})$. Thus, positive adaptivity ($a>0$) increases the effective coupling strength, even if imperfect locking ($\vp^\text{st}$) occurs due to heterogeneity ($\omega \neq 0$). By analogy, a similar argument applies to row-averaged equations \eqref{eq:nodal0}, except that the coupling strength is now modulated with the order parameter. However, in the continuum model negative adaptation does not increase the synchrony domain, while in the two-oscillator case it still increases.
(II) in both cases there exist situations where the oscillators may either lock or drift depending on the initial conditions. However, in the continuum limit for any non-zero adaptation there is a portion of oscillators that can drift or lock, while for two oscillators there is only a specific parameter region that allows where this occurs. (III) In both cases there exists a parameter region which allows (some) oscillators to lock at 2 different phases, and in both cases those regions are for $|a| > 1$. In the continuum limit this allows for the 2-phase clusters, while for two oscillators it just means more possible stable solutions. 
These comparisons highlight the fundamental behavior observed in both systems and demonstrate the general influence of adaptation on phase oscillators, scaling from two oscillators to a continuum.

Future work can investigate how similar our reduced model~\eqref{eq:nodal0} is to the full pairwise model~\eqref{eq:full_system}, better identifying which dynamical features are preserved and which differ between the two. 
One may also extend the considered model in several ways:
(I) Consider a more general phase model, such as an ensemble of adaptive theta neurons and proposed in~\cite{augustsson2024},  which could provide a more biologically relevant perspective.
(II) Generalize the adaptation rule~\eqref{eq:full_system_k} to include, e.g. asymmetric coupling, and see how that affect the coherent patterns, and how one could devise a reduced model similar to~\eqref{eq:nodal0} in that case. 
(III) Examine the effects of noise on the dense multistable states we have identified, showing how noise influences the stability of these states.
Additionally, exploring the memory capacity and learning mechanisms in both the full~\eqref{eq:full_system} and reduced~\eqref{eq:nodal0} models could demonstrate how these systems maintain multiple stable configurations, with potential applications in memory representation and retrieval.

\begin{acknowledgments}
We wish to acknowledge C. Bick and F. Augustsson for helpful comments and discussions.
We gratefully acknowledge financial support from the Royal Swedish Physiographic Society of Lund.
\end{acknowledgments}

\section*{AIP Data Sharing Policy}
Data sharing is not applicable to this article as no new data were created or analyzed in this study.

\appendix

\section{Contribution of drifters to the mean field $R$}\label{sec:drifter_R}
Let us evaluate the integral~\eqref{eq:R_general_integral} just for the drifters, using the expression for their probability density~\eqref{eq:drift_distr}. Importantly, remember that we only consider symmetric frequency distributions $g(\omega) = g(-\omega)$, and the phase density symmetry: $P(\vp+\pi,-\omega) = P(\vp,\omega)$. Combining these two symmetries we can split the frequency integral in two intervals of $\omega$:
\begin{align}
\!\!\!\!\!\!\!\! \int\limits_{|\omega|>\omega_\text{thr}} \!\!\!\!\!\! g(\omega) \int\limits_0^{2\pi} e^{i\vp} P(\vp,\omega) \d\vp\d\omega &= \!\!\!\!\!\!\!\! \int\limits_{\omega < -\omega_\text{thr}} \!\!\!\!\!\! g(\omega) \int\limits_0^{2\pi} e^{i\vp} P(\vp,\omega) \d\vp\d\omega + \nonumber\\ &+ \!\!\!\!\!\! \int\limits_{\omega > \omega_\text{thr}} \!\!\!\!\!\! g(\omega) \int\limits_0^{2\pi} e^{i\vp} P(\vp,\omega) \d\vp\d\omega\,.
\end{align}
In the first integral apply a transformation of variables: $\omega \mapsto -\omega$:
\begin{equation}
\!\!\!\!\!\!\!\! \int\limits_{\omega > \omega_\text{thr}} \!\!\!\!\!\!\!\! g(-\omega) \int\limits_0^{2\pi} e^{i\vp} P(\vp,-\omega) \d\vp\d\omega + \int\limits_{\omega > \omega_\text{thr}} \!\!\!\!\!\!\!\! g(\omega) \int\limits_0^{2\pi} e^{i\vp} P(\vp,\omega) \d\vp\d\omega\,.
\end{equation}
Using the symmetry of the frequency distribution, $g(\omega) = g(-\omega)$, we can combine both terms in a single integral:
\begin{equation}
\int\limits_{\omega > \omega_\text{thr}} g(\omega) \int\limits_0^{2\pi} e^{i\vp} \left[P(\vp,-\omega)+P(\vp,\omega)\right] \d\vp \d\omega\,.
\end{equation}
Recalling the symmetry $P(\vp+\pi,-\omega) = P(\vp,\omega)$, we see that in terms of phase the two contributions (terms of $P$) are simply shifted with respect to each other by a phase-shift of $\pi$, 
\begin{equation}
\int\limits_0^{2\pi} e^{i\vp} \left[P(\vp-\pi,\omega)+P(\vp,\omega) \right] \d\vp\,.
\end{equation}
We  split the integration intervals into two,
\begin{equation}
\int\limits_0^\pi e^{i\vp}\left[ P(\vp-\pi)+P(\vp)\right] \d\vp + \int\limits_\pi^{2\pi} e^{i\vp}\left[P(\vp-\pi)+P(\vp)\right] \d\vp\,,
\end{equation}
and with a transformation of variables: $\vp \mapsto \vp-\pi$ in the second integral (and keeping in mind that $P(\vp)$ is $2\pi$-periodic) we see that the two integrals cancel each other:
\begin{align}
\begin{split}
0&=\int\limits_0^\pi e^{i\vp}\left[P(\vp-\pi)+P(\vp)\right] \d\vp\,, \\
&- \int\limits_0^{\pi} e^{i\vp}\left[P(\vp-2\pi)+P(\vp-\pi)\right] \d\vp\,.\
\end{split}
\end{align}
Therefore, the net contribution of drifters to the mean field $R$ is zero. 

\subsection{Evaluating stationary $\kappa_i$-values for drifters}\label{sec:drifter_ki}
We can express the $\kappa_i^\text{st}$ variables as:
\begin{equation}
\kappa_i^\text{st} = 1+aR\langle \cos(\vp_i) \rangle_t\,.\
\end{equation}
We attempt  to compute the averaged term $\langle \cos(\vp)\rangle_t$. We can rewrite the associated integral in terms of phase $\vp$ using the density~\eqref{eq:drift_distr},
\begin{align}
\langle \cos(\vp) \rangle_t &= \lim_{T\to\infty} \int\limits_0^T \cos(\vp(t))dt \\ &= \frac{1}{2\pi}\int\limits_0^{2\pi}\cos(\vp)P(\vp,\omega)d\vp = \nonumber\\ &= \frac{C}{2\pi}\int\limits_0^{2\pi}\frac{\cos(\vp)}{|\omega-(1+aR\langle \cos(\vp)\rangle_t) R \sin(\vp)|} d\vp\,.
\end{align}
Since the term $\langle \cos(\vp)\rangle_t$ is time-averaged it does not depend on phase $\vp$, meaning that on the right we just have an integral of a trigonometric function over its period, and that equals to zero: $\int\limits_0^{2\pi} \frac{\cos(\vp)}{\omega-A \sin(\vp)}d\vp = 0$, ($A\in\mathbb{R}$ just a constant). We can therefore conclude that the term $\langle \cos(\vp)\rangle_t$ must be zero, 
\begin{equation}
\langle \cos(\vp)\rangle_t = 0\,.
\end{equation}
This result combined with~\eqref{eq:ki_drift_stat} implies that the stationary value for the coupling strengths of drifters simply is given by
\begin{equation}
\kappa_i^\text{st} = 1\,.\
\end{equation}
As a consequence, this means that  -- considering drifters -- the phase equation~\eqref{eq:phase_eq_drift} now looks identical to that of the Kuramoto system~\cite{Kuramoto1975}, and so the same analysis~\cite{strogatz_kuramoto_crawford_2000} applies here.

\section{Iterative solution of the self-consistency equation}\label{sec:iterating_scr}
The fold of the transition for positive $a$ (red line in Fig.~\ref{fig:boundaries}) is solved  numerically by iteratively solving the self-consistency equation~\eqref{eq:scr}:
\begin{widetext}
\begin{subequations}
\begin{align}
f(R) &= \int\limits_{-\vp_\text{thr1}}^{\vp_\text{thr1}} \left[R \cos^2(\vp) + aR^2 \cos(\vp)\cos(2\vp) \right] g\left( R \sin(\vp) + \frac{aR^2}{2} \sin(2\vp) \right) d\vp\,,\\
R_{n+1} &= f(R_n)
\end{align}
\end{subequations}
\end{widetext}
which converges when using an initial guess $0 \leq R_0<1$ and fixed $\sigma$.
We may start this iteration scheme using a large $0<R<1$ as initial value consisten with partial synchrony, as expected for small $\sigma$. Upon convergence, we may progressively increase $\sigma$ and solve for the next value of $R(\sigma)$. This process is continued until we find values of $\sigma$ with $R \to 0$, corresponding to incoherence.

We also know that for high $|a|$ we get the intriguing 2-cluster states. The point at which these may occur is when the phase-frequency relationship ~\eqref{eq:w_to_fi} becomes non-injective, i.e., when
\begin{equation}
R = \frac{1}{|a|}\,.\
\end{equation}
With this condition, we may rewrite the iterative scheme for solving the corresponding self-consistency condition in terms of $a$ only. Thus, we may compute the boundaries of existence for the 2-cluster states (blue lines in Fig.~\ref{fig:boundaries}). First we evaluate the integration boundaries $\vp_\text{thr1}$~\eqref{eq:fithr1}:
\begin{equation}
\vp_\text{\,thr1} = 
\begin{cases}
    \frac{\pi}{3} & \text{for }\ R = \frac{1}{a}\\
    \frac{2\pi}{3}              & \text{for }\ R = -\frac{1}{a}
\end{cases}
\end{equation}
Then we obtain the following iteration scheme:
\begin{widetext}
\begin{subequations}
\begin{align}
f(a) &= \left[ \int\limits_{-\vp_\text{thr1}}^{\vp_\text{thr1}} \left[\frac{1}{a}\cos^2(\vp)+\frac{1}{a}\cos(\vp)\cos(2\vp)\right] g\left(\frac{1}{a}\sin(\vp)+\frac{1}{2a}\sin(2\vp)\right)d\vp \right]^{-1}, \\
a_{n+1} &= f(a_n)\,.\
\end{align}
\end{subequations}
\end{widetext}

\section{Interpretation of expression~\eqref{eq:R_expression_broad} for two-cluster solutions}\label{sec:R_integral}
A crucial way in which the addition of adaptivity generalizes the classical Kuramoto analysis~\cite{Kuramoto1975,strogatz_kuramoto_crawford_2000}, is that the phase-frequency relation~\eqref{eq:w_to_fi} is non-injective. Because of this, expression~\eqref{eq:R_expression_broad}, i.e.
\begin{equation*}
R = \left| \int\limits_{-\omega_\text{thr}}^{\omega_\text{thr}} e^{i\vp(\omega)} g(\omega) \d \omega \right| ,
\end{equation*}
may broadly have to be interpreted as two integrals:
\begin{equation}
R = \left| \int\limits_{-\omega_\text{thr}}^{\omega_\text{thr}} \left[ p e^{i\vp_1(\omega)} + (1-p) e^{i\vp_2(\omega)} \right] g(\omega) \d \omega \right|
\end{equation}
where $\vp_1(\omega)$ and $\vp_2(\omega)$ are the phase relations for the first and second phase cluster respectively, while real-valued $p>0$ and $(1-p)>0$ are their occupations respectively. In general, $p=p(\omega)$ depends on the frequency $\omega$ (and via~\eqref{eq:w_to_fi} on the phase $\vp$), as it encodes different phase configurations between the two phase clusters.  E.g. one could consider a state where
\begin{equation}
p(\omega) = 
\begin{cases}
    0 & \text{for }\ |\omega| < \omega^\dagger\\
    1    & \text{otherwise }
\end{cases}
\end{equation} 
as is the case in Fig.~\ref{fig:boundaries}(c) and Fig.~\ref{fig:numerics}(a). 

\section{Self-consistent equation for the asymmetric phase distribution}\label{sec:asym_distr}
In the main text we only consider symmetric phase distributions, but here we outline how the analysis would look like for the case of states with asymmetric phase distributions. 
We can still move into the co-rotating frame of reference $\Phi = 0$ and still get stationary coupling: $\kappa_i^\text{st} = 1+aR\cos(\varphi_i)$. But the ensemble-averaged dynamic frequency $\langle \dot{\varphi}_i\rangle$ may differ from the mean frequency $\langle \omega_i \rangle$, and is expressed as:
\begin{equation}
\langle \dot{\varphi}_i \rangle = -R \langle \sin(\varphi_i) \rangle - \frac{aR^2}{2} \langle \sin(2\varphi_i) \rangle\,,
\label{eq:meanfreq}
\end{equation}
(we here denote the ensemble average with $\langle \cdot \rangle = \frac{1}{N} \sum_i \cdot $\ ). 
The locked oscillators still all have identical frequency: $\dot{\vp}_i = \langle \vp_i \rangle$ which we use to express the phase-frequency relation:
\begin{equation}
\omega_i = R\left[\sin(\varphi_i)-\langle \sin(\varphi_i) \rangle \right]+\frac{aR^2}{2}\left[\sin(2\varphi_i)-\langle \sin(2\varphi_i) \rangle \right]\,.
\end{equation}
Additionally, since the phase distribution can be asymmetric, we have to be careful about how we evaluate the order parameters:
\begin{equation}
R = \sqrt{\left[ \int \cos(\vp(\omega))g(\omega) \d\omega \right]^2 + \left[ \int \sin(\vp(\omega))g(\omega)\d\omega\right]^2}\,.
\end{equation}
So now the self-consistency iteration has to be done for both $S = \langle \sin(\vp) \rangle$, $S2 = \langle \sin(2 \vp ) \rangle$ and $C = \langle \cos(\vp) \rangle$, with
\begin{widetext}
\begin{subequations}
\begin{align}
S &= \langle \sin(\vp) \rangle = \int \sin(\vp) g(R[\sin(\vp)-S]+\half a R[\sin(2\vp)-S2])(R\sin(\vp) + aR^2 \sin(2\vp)) \d\vp\,,\\
S2 &= \langle \sin(2\vp) \rangle = \int \sin(2\vp) g(R[\sin(\vp)-S]+\half a R[\sin(2\vp)-S2])(R\sin(\vp) + aR^2 \sin(2\vp)) \d\vp\,,\\
C &= \langle \cos(\vp) \rangle = \int \cos(\vp) g(R[\sin(\vp)-S]+\half a R[\sin(2\vp)-S2])(R\sin(\vp) + aR^2 \sin(2\vp)) \d\vp\,.
\end{align}
\end{subequations}
\end{widetext}
Finally, we may then calculate 
\begin{equation}
R = \sqrt{S^2+C^2}.
\end{equation}
This is process is fairly complicated, and we did not see that this iterative approach converges well.  One might also loop through all three values $S,S2,C$ for every $\sigma$, but such a process seems computationally too expensive. For these reasons, we refrained from solving the iterative scheme for asymmetric phase distributions.

\section{Exact analysis of periodic solutions in a simplified model}\label{sec:Duchet}
\begin{figure}[!h]
\begin{overpic}
[width=0.38\textwidth]{"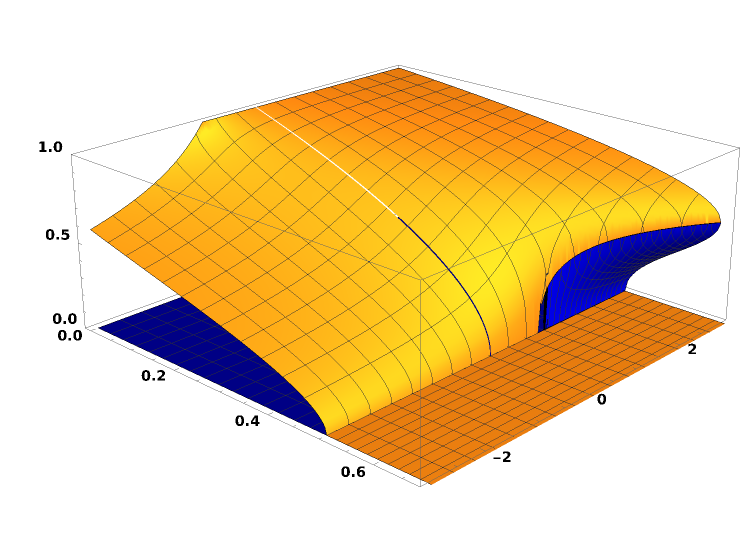"}
\put(84,16){\large\bfseries $a$}
\put(23,15){\large\bfseries $\gamma$}
\put(1,43){\large\bfseries $R$}
\end{overpic}
\caption{Solution branches for $R$  from the simplified Duchet-Bick model \eqref{eq:duchetbick}. }
\label{fig:duchet}
\end{figure}

For studying stability, to see where we can expect periodic solutions, we take an even simpler system, like Duchet {\it et al}~\cite{duchet_bick_2023}.: $\kappa = \frac{1}{N} \sum \kappa_i$:
\begin{subequations}\label{eq:duchetbick}
\begin{align}
\dot{\vp}_i &= \omega_i -\kappa R \sin(\vp_i)\,,\\
\dot{\kappa} &= \epsilon(1+aR^2-\kappa)\,.
\end{align}
\label{eq:duchet}
\end{subequations}
If we consider heterogeneous frequencies drawn from a Cauchy distribution with FWHM $\gamma$, we can further reduce the system to two macroscopic variables $R$ and $\kappa$ (the phase angle of the order parameter trivially decouples):
\begin{subequations}
    \begin{align}
        \dot{R} &= R\left(-\gamma+\frac{\kappa}{2}(1-R^2)\right)\,,\\
        \dot{\kappa} &= \epsilon(1+aR^2-\kappa)\,.
    \end{align}
\label{eq:Duchet}
\end{subequations}
The associated fixed point conditions are
\begin{equation}
    R_0 = \sqrt{1-\frac{2\gamma}{\kappa_0}}\ , \qquad \kappa_0 = 1+aR_0^2\,,
\end{equation}
where we reject the branch with $R_0<0$.
We may iteratively eliminate dependencies on $R_0$ and $\kappa_0$ to obtain explicit parametrizations for the fixed point solutions (+ and - correspond to stable/unstable branches),
\begin{subequations}
\begin{align}\label{eq:duchet_fpbranch_R}
    R_0 &= \sqrt{\frac{a-1 \pm \sqrt{(a+1)^2-8a\gamma}}{2a}}\,, \\
    \label{eq:duchet_fpbranch_kappa}
    \kappa_0 &= \frac{a+1}{2} \pm \sqrt{\left(\frac{a+1}{2}\right)^2-2a\gamma}\,, \
\end{align}
\end{subequations}
where $R_0$ is plotted in Fig.~\ref{fig:duchet}. 
The stable branch for $\gamma = 0$ is $R = 1 \ \text{ if } \ a > -1 ,\ \text{ and else } \ R=1/\sqrt{|a|}$.

The (linear) variational equation is $ \dot{\vec{x}} = {\mathbf J} \cdot \vec{x}$ where $\vec{x} = [ \Delta R, \Delta\kappa ]^T$ with $\Delta R = R-R_0$, $\Delta \kappa =\kappa-\kappa_0$ and the Jacobian 
\begin{equation}
    {\mathbf J} = 
    \begin{bmatrix} 
        -R_0^2\kappa_0 & \gamma \frac{R_0}{\kappa_0} \\ 2a\epsilon R_0 & -\epsilon 
        \end{bmatrix}\,.
\end{equation}
The determinant and trace are
\begin{align}
    \det \, J &= \epsilon R_0^2 \left(\kappa_0-\frac{2\gamma}{\kappa_0}\right)\,, \\ 
    \text{Tr} \, J &= -\epsilon -R_0^2 \kappa_0\,.\
\end{align}
A saddle-node bifurcation between a saddle point and a stable node occurs when $\det\,{J}=0$ and $\text{Tr}\,{J}<0$. The trace remains always negative, since we consider positive values of $\epsilon$ and $R_0$ only, and $\kappa_0$ can also be shown to be positive for all parameters of concern.
The determinant is positive for $a > -1$, and otherwise negative. 
To see this we only have to ask where $\kappa_0-\frac{2\gamma}{\kappa_0}$ changes sign. Applying the fixed point solution \eqref{eq:duchet_fpbranch_kappa} to this condition, we find that crossing the $a = -1$ boundary, the saddle-node bifurcation occurs. As a consequence, the Duchet model does not display non-stationary states above $a=-1$ (see Discussion).


\begin{thebibliography}{27}%
\makeatletter
\providecommand \@ifxundefined [1]{%
 \@ifx{#1\undefined}
}%
\providecommand \@ifnum [1]{%
 \ifnum #1\expandafter \@firstoftwo
 \else \expandafter \@secondoftwo
 \fi
}%
\providecommand \@ifx [1]{%
 \ifx #1\expandafter \@firstoftwo
 \else \expandafter \@secondoftwo
 \fi
}%
\providecommand \natexlab [1]{#1}%
\providecommand \enquote  [1]{``#1''}%
\providecommand \bibnamefont  [1]{#1}%
\providecommand \bibfnamefont [1]{#1}%
\providecommand \citenamefont [1]{#1}%
\providecommand \href@noop [0]{\@secondoftwo}%
\providecommand \href [0]{\begingroup \@sanitize@url \@href}%
\providecommand \@href[1]{\@@startlink{#1}\@@href}%
\providecommand \@@href[1]{\endgroup#1\@@endlink}%
\providecommand \@sanitize@url [0]{\catcode `\\12\catcode `\$12\catcode
  `\&12\catcode `\#12\catcode `\^12\catcode `\_12\catcode `\%12\relax}%
\providecommand \@@startlink[1]{}%
\providecommand \@@endlink[0]{}%
\providecommand \url  [0]{\begingroup\@sanitize@url \@url }%
\providecommand \@url [1]{\endgroup\@href {#1}{\urlprefix }}%
\providecommand \urlprefix  [0]{URL }%
\providecommand \Eprint [0]{\href }%
\providecommand \doibase [0]{http://dx.doi.org/}%
\providecommand \selectlanguage [0]{\@gobble}%
\providecommand \bibinfo  [0]{\@secondoftwo}%
\providecommand \bibfield  [0]{\@secondoftwo}%
\providecommand \translation [1]{[#1]}%
\providecommand \BibitemOpen [0]{}%
\providecommand \bibitemStop [0]{}%
\providecommand \bibitemNoStop [0]{.\EOS\space}%
\providecommand \EOS [0]{\spacefactor3000\relax}%
\providecommand \BibitemShut  [1]{\csname bibitem#1\endcsname}%
\let\auto@bib@innerbib\@empty
\bibitem [{\citenamefont {Pikovsky}, \citenamefont {Rosenblum},\ and\
  \citenamefont {Kurths}(2001)}]{synch_book}%
  \BibitemOpen
  \bibfield  {author} {\bibinfo {author} {\bibfnamefont {A.}~\bibnamefont
  {Pikovsky}}, \bibinfo {author} {\bibfnamefont {M.}~\bibnamefont {Rosenblum}},
  \ and\ \bibinfo {author} {\bibfnamefont {J.}~\bibnamefont {Kurths}},\
  }\href@noop {} {\emph {\bibinfo {title} {Synchronization: A Universal Concept
  in Nonlinear Sciences}}},\ Cambridge Nonlinear Science Series\ (\bibinfo
  {publisher} {Cambridge University Press},\ \bibinfo {year}
  {2001})\BibitemShut {NoStop}%
\bibitem [{\citenamefont {Strogatz}(2003)}]{strogatz_synch}%
  \BibitemOpen
  \bibfield  {author} {\bibinfo {author} {\bibfnamefont {S.~H.}\ \bibnamefont
  {Strogatz}},\ }\href@noop {} {\emph {\bibinfo {title} {Sync: The Emerging
  Science of Spontaneous Order}}}\ (\bibinfo  {publisher} {Hachette Books},\
  \bibinfo {year} {2003})\BibitemShut {NoStop}%
\bibitem [{hea()}]{heart_example}%
  \BibitemOpen
  \bibfield  {title} {\enquote {\bibinfo {title} {Mechanisms of sinoatrial
  pacemaker synchronization: a new hypothesis.}}\ }\href {\doibase
  10.1161/01.RES.61.5.704} {\bibfield  {journal} {\bibinfo  {journal} {Circ.
  Res.}\ }\textbf {\bibinfo {volume} {61}},\
  10.1161/01.RES.61.5.704}\BibitemShut {NoStop}%
\bibitem [{\citenamefont {Rohden}\ \emph {et~al.}(2012)\citenamefont {Rohden},
  \citenamefont {Sorge}, \citenamefont {Timme}, \citenamefont {Timme},\ and\
  \citenamefont {Witthaut}}]{power_grids_example}%
  \BibitemOpen
  \bibfield  {author} {\bibinfo {author} {\bibfnamefont {M.}~\bibnamefont
  {Rohden}}, \bibinfo {author} {\bibfnamefont {A.}~\bibnamefont {Sorge}},
  \bibinfo {author} {\bibfnamefont {M.}~\bibnamefont {Timme}}, \bibinfo
  {author} {\bibfnamefont {M.}~\bibnamefont {Timme}}, \ and\ \bibinfo {author}
  {\bibfnamefont {D.}~\bibnamefont {Witthaut}},\ }\bibfield  {title} {\enquote
  {\bibinfo {title} {Self-organized synchronization in decentralized power
  grids},}\ }\href {\doibase 10.1103/physrevlett.109.064101} {\ \textbf
  {\bibinfo {volume} {109}} (\bibinfo {year} {2012}),\
  10.1103/physrevlett.109.064101}\BibitemShut {NoStop}%
\bibitem [{\citenamefont {Singer}\ and\ \citenamefont
  {Gray}(1995)}]{neural_networks_example1}%
  \BibitemOpen
  \bibfield  {author} {\bibinfo {author} {\bibfnamefont {W.}~\bibnamefont
  {Singer}}\ and\ \bibinfo {author} {\bibfnamefont {C.~M.}\ \bibnamefont
  {Gray}},\ }\bibfield  {title} {\enquote {\bibinfo {title} {Visual feature
  integration and the temporal correlation hypothesis},}\ }\href {\doibase
  10.1146/ANNUREV.NE.18.030195.003011} {\ \textbf {\bibinfo {volume} {18}}
  (\bibinfo {year} {1995}),\ 10.1146/ANNUREV.NE.18.030195.003011}\BibitemShut
  {NoStop}%
\bibitem [{\citenamefont {Uhlhaas}\ \emph {et~al.}(2006)\citenamefont
  {Uhlhaas}, \citenamefont {Uhlhaas}, \citenamefont {Singer},\ and\
  \citenamefont {Singer}}]{neural_networks_example2}%
  \BibitemOpen
  \bibfield  {author} {\bibinfo {author} {\bibfnamefont {P.~J.}\ \bibnamefont
  {Uhlhaas}}, \bibinfo {author} {\bibfnamefont {P.~J.}\ \bibnamefont
  {Uhlhaas}}, \bibinfo {author} {\bibfnamefont {W.}~\bibnamefont {Singer}}, \
  and\ \bibinfo {author} {\bibfnamefont {W.}~\bibnamefont {Singer}},\
  }\bibfield  {title} {\enquote {\bibinfo {title} {Neural synchrony in brain
  disorders: Relevance for cognitive dysfunctions and pathophysiology},}\
  }\href {\doibase 10.1016/J.NEURON.2006.09.020} {\ \textbf {\bibinfo {volume}
  {52}} (\bibinfo {year} {2006}),\ 10.1016/J.NEURON.2006.09.020}\BibitemShut
  {NoStop}%
\bibitem [{\citenamefont {Seliger}, \citenamefont {Young},\ and\ \citenamefont
  {Tsimring}(2002)}]{adaptive_example1}%
  \BibitemOpen
  \bibfield  {author} {\bibinfo {author} {\bibfnamefont {P.}~\bibnamefont
  {Seliger}}, \bibinfo {author} {\bibfnamefont {S.~C.}\ \bibnamefont {Young}},
  \ and\ \bibinfo {author} {\bibfnamefont {L.~S.}\ \bibnamefont {Tsimring}},\
  }\bibfield  {title} {\enquote {\bibinfo {title} {Plasticity and learning in a
  network of coupled phase oscillators},}\ }\href {\doibase
  10.1103/PHYSREVE.65.041906} {\ \textbf {\bibinfo {volume} {65}} (\bibinfo
  {year} {2002}),\ 10.1103/PHYSREVE.65.041906}\BibitemShut {NoStop}%
\bibitem [{\citenamefont {Maistrenko}\ \emph {et~al.}(2007)\citenamefont
  {Maistrenko}, \citenamefont {Lysyansky}, \citenamefont {Hauptmann},
  \citenamefont {Burylko},\ and\ \citenamefont {Tass}}]{adaptive_example2}%
  \BibitemOpen
  \bibfield  {author} {\bibinfo {author} {\bibfnamefont {Y.~L.}\ \bibnamefont
  {Maistrenko}}, \bibinfo {author} {\bibfnamefont {B.}~\bibnamefont
  {Lysyansky}}, \bibinfo {author} {\bibfnamefont {C.}~\bibnamefont
  {Hauptmann}}, \bibinfo {author} {\bibfnamefont {O.}~\bibnamefont {Burylko}},
  \ and\ \bibinfo {author} {\bibfnamefont {P.~A.}\ \bibnamefont {Tass}},\
  }\bibfield  {title} {\enquote {\bibinfo {title} {Multistability in the
  kuramoto model with synaptic plasticity},}\ }\href {\doibase
  10.1103/PhysRevE.75.066207} {\bibfield  {journal} {\bibinfo  {journal} {Phys.
  Rev. E}\ }\textbf {\bibinfo {volume} {75}},\ \bibinfo {pages} {066207}
  (\bibinfo {year} {2007})}\BibitemShut {NoStop}%
\bibitem [{\citenamefont {Martens}\ and\ \citenamefont
  {Klemm}(2017)}]{vascular_example}%
  \BibitemOpen
  \bibfield  {author} {\bibinfo {author} {\bibfnamefont {E.~A.}\ \bibnamefont
  {Martens}}\ and\ \bibinfo {author} {\bibfnamefont {K.}~\bibnamefont
  {Klemm}},\ }\bibfield  {title} {\enquote {\bibinfo {title} {Transitions from
  trees to cycles in adaptive flow networks},}\ }\href {\doibase
  10.3389/fphy.2017.00062} {\bibfield  {journal} {\bibinfo  {journal}
  {Frontiers in Physics}\ }\textbf {\bibinfo {volume} {5}} (\bibinfo {year}
  {2017}),\ 10.3389/fphy.2017.00062}\BibitemShut {NoStop}%
\bibitem [{\citenamefont {Mestre}\ \emph {et~al.}(2020)\citenamefont {Mestre},
  \citenamefont {Du}, \citenamefont {Sweeney}, \citenamefont {Liu},
  \citenamefont {Samson}, \citenamefont {Peng}, \citenamefont {Mortensen},
  \citenamefont {Stæger}, \citenamefont {Bork}, \citenamefont {Bashford},
  \citenamefont {Toro}, \citenamefont {Tithof}, \citenamefont {Kelley},
  \citenamefont {Thomas}, \citenamefont {Hjorth}, \citenamefont {Martens},
  \citenamefont {Mehta}, \citenamefont {Solis}, \citenamefont {Blinder},
  \citenamefont {Kleinfeld}, \citenamefont {Hirase}, \citenamefont {Mori},\
  and\ \citenamefont {Nedergaard}}]{glymphatic_example1}%
  \BibitemOpen
  \bibfield  {author} {\bibinfo {author} {\bibfnamefont {H.}~\bibnamefont
  {Mestre}}, \bibinfo {author} {\bibfnamefont {T.}~\bibnamefont {Du}}, \bibinfo
  {author} {\bibfnamefont {A.~M.}\ \bibnamefont {Sweeney}}, \bibinfo {author}
  {\bibfnamefont {G.}~\bibnamefont {Liu}}, \bibinfo {author} {\bibfnamefont
  {A.~J.}\ \bibnamefont {Samson}}, \bibinfo {author} {\bibfnamefont
  {W.}~\bibnamefont {Peng}}, \bibinfo {author} {\bibfnamefont {K.~N.}\
  \bibnamefont {Mortensen}}, \bibinfo {author} {\bibfnamefont {F.~F.}\
  \bibnamefont {Stæger}}, \bibinfo {author} {\bibfnamefont {P.~A.~R.}\
  \bibnamefont {Bork}}, \bibinfo {author} {\bibfnamefont {L.}~\bibnamefont
  {Bashford}}, \bibinfo {author} {\bibfnamefont {E.~R.}\ \bibnamefont {Toro}},
  \bibinfo {author} {\bibfnamefont {J.}~\bibnamefont {Tithof}}, \bibinfo
  {author} {\bibfnamefont {D.~H.}\ \bibnamefont {Kelley}}, \bibinfo {author}
  {\bibfnamefont {J.~H.}\ \bibnamefont {Thomas}}, \bibinfo {author}
  {\bibfnamefont {P.~G.}\ \bibnamefont {Hjorth}}, \bibinfo {author}
  {\bibfnamefont {E.~A.}\ \bibnamefont {Martens}}, \bibinfo {author}
  {\bibfnamefont {R.~I.}\ \bibnamefont {Mehta}}, \bibinfo {author}
  {\bibfnamefont {O.}~\bibnamefont {Solis}}, \bibinfo {author} {\bibfnamefont
  {P.}~\bibnamefont {Blinder}}, \bibinfo {author} {\bibfnamefont
  {D.}~\bibnamefont {Kleinfeld}}, \bibinfo {author} {\bibfnamefont
  {H.}~\bibnamefont {Hirase}}, \bibinfo {author} {\bibfnamefont
  {Y.}~\bibnamefont {Mori}}, \ and\ \bibinfo {author} {\bibfnamefont
  {M.}~\bibnamefont {Nedergaard}},\ }\bibfield  {title} {\enquote {\bibinfo
  {title} {Cerebrospinal fluid influx drives acute ischemic tissue swelling},}\
  }\href {\doibase 10.1126/science.aax7171} {\bibfield  {journal} {\bibinfo
  {journal} {Science}\ }\textbf {\bibinfo {volume} {367}},\ \bibinfo {pages}
  {eaax7171} (\bibinfo {year} {2020})},\ \Eprint
  {http://arxiv.org/abs/https://www.science.org/doi/pdf/10.1126/science.aax7171}
  {https://www.science.org/doi/pdf/10.1126/science.aax7171} \BibitemShut
  {NoStop}%
\bibitem [{\citenamefont {Taylor-King}\ \emph {et~al.}(2017)\citenamefont
  {Taylor-King}, \citenamefont {Basanta}, \citenamefont {Chapman},\ and\
  \citenamefont {Porter}}]{osteocyte_example}%
  \BibitemOpen
  \bibfield  {author} {\bibinfo {author} {\bibfnamefont {J.~P.}\ \bibnamefont
  {Taylor-King}}, \bibinfo {author} {\bibfnamefont {D.}~\bibnamefont
  {Basanta}}, \bibinfo {author} {\bibfnamefont {S.~J.}\ \bibnamefont
  {Chapman}}, \ and\ \bibinfo {author} {\bibfnamefont {M.~A.}\ \bibnamefont
  {Porter}},\ }\bibfield  {title} {\enquote {\bibinfo {title} {Mean-field
  approach to evolving spatial networks, with an application to osteocyte
  network formation},}\ }\href {\doibase 10.1103/PhysRevE.96.012301} {\bibfield
   {journal} {\bibinfo  {journal} {Phys. Rev. E}\ }\textbf {\bibinfo {volume}
  {96}},\ \bibinfo {pages} {012301} (\bibinfo {year} {2017})}\BibitemShut
  {NoStop}%
\bibitem [{\citenamefont {Skyrms}\ and\ \citenamefont
  {Pemantle}(2009)}]{social_example}%
  \BibitemOpen
  \bibfield  {author} {\bibinfo {author} {\bibfnamefont {B.}~\bibnamefont
  {Skyrms}}\ and\ \bibinfo {author} {\bibfnamefont {R.}~\bibnamefont
  {Pemantle}},\ }\enquote {\bibinfo {title} {A dynamic model of social network
  formation},}\ in\ \href {\doibase 10.1007/978-3-642-01284-6_11} {\emph
  {\bibinfo {booktitle} {Adaptive Networks: Theory, Models and
  Applications}}},\ \bibinfo {editor} {edited by\ \bibinfo {editor}
  {\bibfnamefont {T.}~\bibnamefont {Gross}}\ and\ \bibinfo {editor}
  {\bibfnamefont {H.}~\bibnamefont {Sayama}}}\ (\bibinfo  {publisher} {Springer
  Berlin Heidelberg},\ \bibinfo {address} {Berlin, Heidelberg},\ \bibinfo
  {year} {2009})\ pp.\ \bibinfo {pages} {231--251}\BibitemShut {NoStop}%
\bibitem [{\citenamefont {Gerstner}\ and\ \citenamefont
  {Kistler}(2002)}]{neural_net_example}%
  \BibitemOpen
  \bibfield  {author} {\bibinfo {author} {\bibfnamefont {W.}~\bibnamefont
  {Gerstner}}\ and\ \bibinfo {author} {\bibfnamefont {W.~M.}\ \bibnamefont
  {Kistler}},\ }\bibfield  {title} {\enquote {\bibinfo {title} {Mathematical
  formulations of hebbian learning},}\ }\href {\doibase
  10.1007/S00422-002-0353-Y} {\bibfield  {journal} {\bibinfo  {journal} {Biol.
  Cybern.}\ }\textbf {\bibinfo {volume} {87}} (\bibinfo {year} {2002}),\
  10.1007/S00422-002-0353-Y}\BibitemShut {NoStop}%
\bibitem [{\citenamefont {Jüttner}\ and\ \citenamefont
  {Martens}(2023)}]{juttner_martens_2023}%
  \BibitemOpen
  \bibfield  {author} {\bibinfo {author} {\bibfnamefont {B.}~\bibnamefont
  {Jüttner}}\ and\ \bibinfo {author} {\bibfnamefont {E.~A.}\ \bibnamefont
  {Martens}},\ }\bibfield  {title} {\enquote {\bibinfo {title} {{Complex
  dynamics in adaptive phase oscillator networks}},}\ }\href {\doibase
  10.1063/5.0133190} {\bibfield  {journal} {\bibinfo  {journal} {Chaos: An
  Interdisciplinary Journal of Nonlinear Science}\ }\textbf {\bibinfo {volume}
  {33}},\ \bibinfo {pages} {053106} (\bibinfo {year} {2023})}\BibitemShut
  {NoStop}%
\bibitem [{\citenamefont {Aoki}\ and\ \citenamefont
  {Aoyagi}(2009)}]{aoki_aoyagi_2009}%
  \BibitemOpen
  \bibfield  {author} {\bibinfo {author} {\bibfnamefont {T.}~\bibnamefont
  {Aoki}}\ and\ \bibinfo {author} {\bibfnamefont {T.}~\bibnamefont {Aoyagi}},\
  }\bibfield  {title} {\enquote {\bibinfo {title} {Co-evolution of phases and
  connection strengths in a network of phase oscillators},}\ }\href {\doibase
  10.1103/PhysRevLett.102.034101} {\bibfield  {journal} {\bibinfo  {journal}
  {Phys. Rev. Lett.}\ }\textbf {\bibinfo {volume} {102}},\ \bibinfo {pages}
  {034101} (\bibinfo {year} {2009})}\BibitemShut {NoStop}%
\bibitem [{\citenamefont {Kuramoto}(1975)}]{Kuramoto1975}%
  \BibitemOpen
  \bibfield  {author} {\bibinfo {author} {\bibfnamefont {Y.}~\bibnamefont
  {Kuramoto}},\ }\bibfield  {title} {\enquote {\bibinfo {title}
  {Self-entrainment of a population of coupled non-linear oscillators},}\
  }\href {https://api.semanticscholar.org/CorpusID:53832482} {\bibfield
  {journal} {\bibinfo  {journal} {In International Symposium on Mathematical
  Problems in Theoretical Physics}\ } (\bibinfo {year} {1975})}\BibitemShut
  {NoStop}%
\bibitem [{\citenamefont {Strogatz}(2000)}]{strogatz_kuramoto_crawford_2000}%
  \BibitemOpen
  \bibfield  {author} {\bibinfo {author} {\bibfnamefont {S.~H.}\ \bibnamefont
  {Strogatz}},\ }\bibfield  {title} {\enquote {\bibinfo {title} {{From Kuramoto
  to Crawford: exploring the onset of synchronization in populations of coupled
  oscillators}},}\ }\href {\doibase
  https://doi.org/10.1016/S0167-2789(00)00094-4} {\bibfield  {journal}
  {\bibinfo  {journal} {Physica D: Nonlinear Phenomena}\ }\textbf {\bibinfo
  {volume} {143}},\ \bibinfo {pages} {1--20} (\bibinfo {year}
  {2000})}\BibitemShut {NoStop}%
\bibitem [{\citenamefont {Martens}\ \emph {et~al.}(2009)\citenamefont
  {Martens}, \citenamefont {Barreto}, \citenamefont {Strogatz}, \citenamefont
  {Ott}, \citenamefont {So},\ and\ \citenamefont
  {Antonsen}}]{martens2009exact}%
  \BibitemOpen
  \bibfield  {author} {\bibinfo {author} {\bibfnamefont {E.~A.}\ \bibnamefont
  {Martens}}, \bibinfo {author} {\bibfnamefont {E.}~\bibnamefont {Barreto}},
  \bibinfo {author} {\bibfnamefont {S.~H.}\ \bibnamefont {Strogatz}}, \bibinfo
  {author} {\bibfnamefont {E.}~\bibnamefont {Ott}}, \bibinfo {author}
  {\bibfnamefont {P.}~\bibnamefont {So}}, \ and\ \bibinfo {author}
  {\bibfnamefont {T.~M.}\ \bibnamefont {Antonsen}},\ }\bibfield  {title}
  {\enquote {\bibinfo {title} {{Exact results for the Kuramoto model with a
  bimodal frequency distribution}},}\ }\href@noop {} {\bibfield  {journal}
  {\bibinfo  {journal} {Physical Review E—Statistical, Nonlinear, and Soft
  Matter Physics}\ }\textbf {\bibinfo {volume} {79}},\ \bibinfo {pages}
  {026204} (\bibinfo {year} {2009})}\BibitemShut {NoStop}%
\bibitem [{\citenamefont {Thiele}\ \emph {et~al.}(2023)\citenamefont {Thiele},
  \citenamefont {Berner}, \citenamefont {Tass}, \citenamefont {Schöll},\ and\
  \citenamefont {Yanchuk}}]{recurrent_synchronization}%
  \BibitemOpen
  \bibfield  {author} {\bibinfo {author} {\bibfnamefont {M.}~\bibnamefont
  {Thiele}}, \bibinfo {author} {\bibfnamefont {R.}~\bibnamefont {Berner}},
  \bibinfo {author} {\bibfnamefont {P.~A.}\ \bibnamefont {Tass}}, \bibinfo
  {author} {\bibfnamefont {E.}~\bibnamefont {Schöll}}, \ and\ \bibinfo
  {author} {\bibfnamefont {S.}~\bibnamefont {Yanchuk}},\ }\bibfield  {title}
  {\enquote {\bibinfo {title} {Asymmetric adaptivity induces recurrent
  synchronization in complex networks},}\ }\href {\doibase 10.1063/5.0128102}
  {\bibfield  {journal} {\bibinfo  {journal} {Chaos: An Interdisciplinary
  Journal of Nonlinear Science}\ }\textbf {\bibinfo {volume} {33}},\ \bibinfo
  {pages} {023123} (\bibinfo {year} {2023})}\BibitemShut {NoStop}%
\bibitem [{\citenamefont {Sales}, \citenamefont {Yanchuk},\ and\ \citenamefont
  {Kurths}(2024)}]{chaotic_recurrent_synchronization_arxiv}%
  \BibitemOpen
  \bibfield  {author} {\bibinfo {author} {\bibfnamefont {M.~R.}\ \bibnamefont
  {Sales}}, \bibinfo {author} {\bibfnamefont {S.}~\bibnamefont {Yanchuk}}, \
  and\ \bibinfo {author} {\bibfnamefont {J.}~\bibnamefont {Kurths}},\
  }\bibfield  {title} {\enquote {\bibinfo {title} {Recurrent chaotic clustering
  and slow chaos in adaptive networks},}\ }\href@noop {} {\  (\bibinfo {year}
  {2024})},\ \Eprint {http://arxiv.org/abs/2402.17646} {arXiv:2402.17646}
  \BibitemShut {NoStop}%
\bibitem [{\citenamefont {Duchet}, \citenamefont {Bick},\ and\ \citenamefont
  {Byrne}(2023)}]{duchet_bick_2023}%
  \BibitemOpen
  \bibfield  {author} {\bibinfo {author} {\bibfnamefont {B.}~\bibnamefont
  {Duchet}}, \bibinfo {author} {\bibfnamefont {C.}~\bibnamefont {Bick}}, \ and\
  \bibinfo {author} {\bibfnamefont {{\'A}.}~\bibnamefont {Byrne}},\ }\bibfield
  {title} {\enquote {\bibinfo {title} {Mean-field approximations with adaptive
  coupling for networks with spike-timing-dependent plasticity},}\ }\href
  {\doibase 10.1162/neco_a_01601} {\ \textbf {\bibinfo {volume} {35}},\
  \bibinfo {pages} {1481--1528} (\bibinfo {year} {2023})}\BibitemShut {NoStop}%
\bibitem [{\citenamefont {Guckenheimer}\ and\ \citenamefont
  {Holmes}(2013)}]{guckenheimer2013nonlinear}%
  \BibitemOpen
  \bibfield  {author} {\bibinfo {author} {\bibfnamefont {J.}~\bibnamefont
  {Guckenheimer}}\ and\ \bibinfo {author} {\bibfnamefont {P.}~\bibnamefont
  {Holmes}},\ }\href@noop {} {\emph {\bibinfo {title} {Nonlinear oscillations,
  dynamical systems, and bifurcations of vector fields}}},\ Vol.~\bibinfo
  {volume} {42}\ (\bibinfo  {publisher} {Springer Science \& Business Media},\
  \bibinfo {year} {2013})\BibitemShut {NoStop}%
\bibitem [{\citenamefont {Kuehn}\ \emph {et~al.}(2015)\citenamefont {Kuehn}
  \emph {et~al.}}]{kuehn2015multiple}%
  \BibitemOpen
  \bibfield  {author} {\bibinfo {author} {\bibfnamefont {C.}~\bibnamefont
  {Kuehn}} \emph {et~al.},\ }\href@noop {} {\emph {\bibinfo {title} {Multiple
  time scale dynamics}}},\ Vol.\ \bibinfo {volume} {191}\ (\bibinfo
  {publisher} {Springer},\ \bibinfo {year} {2015})\BibitemShut {NoStop}%
\bibitem [{\citenamefont {Berner}\ \emph {et~al.}(2019)\citenamefont {Berner},
  \citenamefont {Fialkowski}, \citenamefont {Kasatkin}, \citenamefont
  {Nekorkin}, \citenamefont {Yanchuk},\ and\ \citenamefont
  {Schöll}}]{10.1063/1.5097835}%
  \BibitemOpen
  \bibfield  {author} {\bibinfo {author} {\bibfnamefont {R.}~\bibnamefont
  {Berner}}, \bibinfo {author} {\bibfnamefont {J.}~\bibnamefont {Fialkowski}},
  \bibinfo {author} {\bibfnamefont {D.}~\bibnamefont {Kasatkin}}, \bibinfo
  {author} {\bibfnamefont {V.}~\bibnamefont {Nekorkin}}, \bibinfo {author}
  {\bibfnamefont {S.}~\bibnamefont {Yanchuk}}, \ and\ \bibinfo {author}
  {\bibfnamefont {E.}~\bibnamefont {Schöll}},\ }\bibfield  {title} {\enquote
  {\bibinfo {title} {{Hierarchical frequency clusters in adaptive networks of
  phase oscillators}},}\ }\href {\doibase 10.1063/1.5097835} {\bibfield
  {journal} {\bibinfo  {journal} {Chaos: An Interdisciplinary Journal of
  Nonlinear Science}\ }\textbf {\bibinfo {volume} {29}},\ \bibinfo {pages}
  {103134} (\bibinfo {year} {2019})},\ \Eprint
  {http://arxiv.org/abs/https://pubs.aip.org/aip/cha/article-pdf/doi/10.1063/1.5097835/14624675/103134\_1\_online.pdf}
  {https://pubs.aip.org/aip/cha/article-pdf/doi/10.1063/1.5097835/14624675/103134\_1\_online.pdf}
  \BibitemShut {NoStop}%
\bibitem [{\citenamefont {Ott}\ and\ \citenamefont
  {Antonsen}(2008)}]{OttAntonsen}%
  \BibitemOpen
  \bibfield  {author} {\bibinfo {author} {\bibfnamefont {E.}~\bibnamefont
  {Ott}}\ and\ \bibinfo {author} {\bibfnamefont {T.~M.}\ \bibnamefont
  {Antonsen}},\ }\bibfield  {title} {\enquote {\bibinfo {title} {{Low
  dimensional behavior of large systems of globally coupled oscillators}},}\
  }\href {\doibase 10.1063/1.2930766} {\bibfield  {journal} {\bibinfo
  {journal} {Chaos: An Interdisciplinary Journal of Nonlinear Science}\
  }\textbf {\bibinfo {volume} {18}},\ \bibinfo {pages} {037113} (\bibinfo
  {year} {2008})}\BibitemShut {NoStop}%
\bibitem [{\citenamefont {Bick}\ \emph {et~al.}(2020)\citenamefont {Bick},
  \citenamefont {Goodfellow}, \citenamefont {Laing},\ and\ \citenamefont
  {Martens}}]{bick2020}%
  \BibitemOpen
  \bibfield  {author} {\bibinfo {author} {\bibfnamefont {C.}~\bibnamefont
  {Bick}}, \bibinfo {author} {\bibfnamefont {M.}~\bibnamefont {Goodfellow}},
  \bibinfo {author} {\bibfnamefont {C.~R.}\ \bibnamefont {Laing}}, \ and\
  \bibinfo {author} {\bibfnamefont {E.~A.}\ \bibnamefont {Martens}},\
  }\bibfield  {title} {\enquote {\bibinfo {title} {Understanding the dynamics
  of biological and neural oscillator networks through exact mean-field
  reductions: a review},}\ }\href {\doibase 10.1186/S13408-020-00086-9}
  {\bibfield  {journal} {\bibinfo  {journal} {Journal of Mathematical
  Neuroscience}\ }\textbf {\bibinfo {volume} {10}} (\bibinfo {year} {2020}),\
  10.1186/S13408-020-00086-9}\BibitemShut {NoStop}%
\bibitem [{\citenamefont {Augustsson}\ and\ \citenamefont
  {Martens}(2024)}]{augustsson2024}%
  \BibitemOpen
  \bibfield  {author} {\bibinfo {author} {\bibfnamefont {F.}~\bibnamefont
  {Augustsson}}\ and\ \bibinfo {author} {\bibfnamefont {E.~A.}\ \bibnamefont
  {Martens}},\ }\href {https://arxiv.org/abs/2407.01089} {\enquote {\bibinfo
  {title} {Co-evolutionary dynamics for two adaptively coupled theta
  neurons},}\ } (\bibinfo {year} {2024}),\ \Eprint
  {http://arxiv.org/abs/2407.01089} {arXiv:2407.01089 [nlin.AO]} \BibitemShut
  {NoStop}%
\end{thebibliography}
%

\end{document}